\documentclass[aps,pra,twocolumn,floatfix,superscriptaddress,times,10pt,showkeys]{revtex4-2}
\usepackage{mathbbol}
\usepackage{amsmath,amssymb}
\usepackage{tikz}
\usepackage{centernot}
\usepackage[utf8]{inputenc}
\usepackage[english]{babel}
\usepackage{amsfonts}
\usepackage{amsthm}
\usepackage{mathrsfs}
\usepackage{mathtools}
\usepackage[nice]{nicefrac}
\usepackage{xfrac}
\usepackage{tabularx}
\usepackage{physics,subfigure}
\usepackage{tikz}
\usetikzlibrary{positioning}
\usepackage{tkz-berge}
\usetikzlibrary{arrows}
\usetikzlibrary{graphs,graphs.standard,fit,shapes.geometric,calc,hobby}
\usepackage[colorlinks=true,hyperfootnotes=true,breaklinks=true,citecolor=red,urlcolor=blue,linkcolor=blue]{hyperref}

\usepackage{array}
\newcolumntype{C}[1]{>{\centering\let\newline\\\arraybackslash\hspace{0pt}}m{#1}}

\begin{document}
\title{From spectral structure to sensing limits in quantum thermometry}
\author{Youssef Aiache}
\email{youssefaiache0@gmail.com}
\affiliation{Laboratory of R\&D in Engineering Sciences, Faculty of Sciences and Techniques Al-Hoceima, Abdelmalek Essaadi University, BP 34. Ajdir 32003, Tetouan, Morocco}
\author{Simone Cavazzoni}
\email{simone.cavazzoni@unimore.it}
\affiliation{Dipartimento di Scienze Fisiche, Informatiche e Matematiche, Universit\`{a} di Modena e Reggio Emilia, I-41125 Modena, Italy}
\author{Abderrahim El Allati}
\email{eabderrahim@uae.ac.ma}
\affiliation{Laboratory of R\&D in Engineering Sciences, Faculty of Sciences and Techniques Al-Hoceima, Abdelmalek Essaadi University, BP 34. Ajdir 32003, Tetouan, Morocco}
\author{Paolo Bordone}
\email{paolo.bordone@unimore.it}
\affiliation{Dipartimento di Scienze Fisiche, Informatiche e Matematiche, Universit\`{a} di Modena e Reggio Emilia, I-41125 Modena, Italy}
\affiliation{Centro S3, CNR-Istituto di Nanoscienze, I-41125 Modena, Italy}
\author{Matteo G. A. Paris}
\email{matteo.paris@fisica.unimi.it}
\affiliation{Dipartimento di Fisica dell'Universit\`{a} di Milano, I-20133 Milano, Italy}
\date{\today}
\begin{abstract}
The precision of a quantum thermometer is fundamentally constrained by the spectral structure of the probe itself, and a systematic mapping between energy-level configurations and thermometric performance provides relevant information for the design of optimized devices. In this work, we establish such a mapping by analyzing a broad class of quantum systems, ranging from finite spin ensembles and degenerate atoms to confining potentials, quantum walks, and continuous-spectrum models. 
We derive asymptotic laws for the quantum Fisher information (QFI) and show that, at high temperature, bounded spectra yield a $T^{-4}$ decay, while scale-free unbounded spectra yield a slower $T^{-2}$ decay. At low temperatures, we show that sensitivity, though universally exponentially suppressed, can be enhanced by engineering a degenerate first excited level, e.g., through a quantum walk on a fully connected topology. By contrast, specific quantum walk topologies provide a distinct enhancement mechanism based on gap engineering, whereby specific conditions produce a $T^{-2}$ low-temperature scaling. 
Furthermore, we also derive exact formulas and asymptotic scaling for any power-law or hybrid spectrum, showing general theoretical properties of quantum thermometry. Our results contribute to the transformation of spectral information into a resource for quantum thermometry, providing both fundamental bounds and practical guidelines for tailored temperature sensing.
\end{abstract}
\keywords{Quantum thermometry, Quantum estimation, Quantum sensing}
\maketitle
\section{Introduction}
\label{sec:I}
Quantum thermometry plays a central role in estimation theory, not only because temperature is a fundamental parameter in virtually every physical setting but also because thermal effects pose persistent challenges to the control and reliability of quantum devices. Unknown temperature fluctuations can induce errors in quantum computing protocols \cite{Nielsen_Chuang_2010,riera2012thermalization,basak2026noise}, degrade the efficiency of quantum engines, batteries, and refrigerators \cite{callies2023structuring,cavazzoni2026quantum,Ye2026Measuring}, and limit the reproducibility of experimental results \cite{daniel2010new,moore2021effect}. Most experiments operate at finite temperature, where the quantum system of interest is weakly coupled to an external reservoir and relaxes to a thermal state. In this general regime, estimating the environmental temperature is not merely a calibration step but rather a central part of characterizing any equilibrium quantum system. 

Therefore, the development of optimized quantum thermometers is crucial \cite{tham2016simulating,xie2017optimal,mukherjee2019enhanced,pati2020quantum,KulikovPhysRevLett2020,hovhannisyan2021optimal,planella2022bath,ullah2023low,abiuso2024optimal,AiachePhysRevA2024,LvovPhysRevApplied2025,AiachePRAQM2025}. Quantum estimation theory provides the fundamental tools to saturate the ultimate precision bounds on temperature measurements \cite{miller2018energy,mehboudi2019thermometry}, and recent years have seen several proposals for experimental realizations \cite{correa2015individual,mancino2017quantum,seah2019collisional,xie2020quantum,montenegro2020mechanical,mitchison2020situ,o2021stochastic,fujiwara2021diamond,zhang2022approaching,dedyulin2022emerging,briant2022photonic,yang2024sequential,liu2024all,AiachePhysRevE2024,mehboudi2025optimal,akamatsu2025fundamental}. For equilibrium sensing, which is the focus of this work, it is well known that measuring the probe energy saturates the quantum limit \cite{StacePhysRevA2010,LatuneNJP2020,Glatthard2023energymeasurements,ChangPhysRevRes2024}, regardless of temperature. However, this optimality of the energy measurements does not imply universal performance across different probes. Depending on their spectral structure or critical behavior, different quantum systems can exhibit markedly different sensitivities to thermal fluctuations \cite{Campbell2018QST,Salado_Mejia_2021,yuan2023quantum,UllahPhysRevA2025}. This immediately raises two questions: what spectral features characterize an optimal probe for equilibrium quantum thermometry? And how does this optimality depend on the underlying energy spectrum \cite{mok2021optimal,potts2019fundamental}?

Answering these questions requires a systematic analysis of how temperature information is encoded in the equilibrium states of different classes of quantum systems. The distribution of energy levels, the presence and structure of degeneracies, the scaling of spectral gaps, and the distinction between bounded and unbounded spectra jointly determine the low- and high-temperature sensitivity of a quantum thermometer. From this point of view, the energy spectrum itself becomes a thermodynamic resource \cite{zhang2024temperature}, as it dictates not only the ultimate achievable precision but also the accessible temperature range and how performance scales with system size.

In this work, we present a systematic spectral analysis of equilibrium quantum thermometry. We consider a broad class of probes, ranging from finite-dimensional spins and degenerate atoms to confining potentials, quantum walks, hybrid discrete-continuous spectra, atomic and molecular systems, and free particles. We analyze how level spacing, degeneracy, spectral growth, and spectral boundedness affect the quantum Fisher information (QFI) for temperature estimation. Special emphasis is placed on the low- and high-temperature regimes, where the QFI admits analytical asymptotic expressions. \textcolor{black}{Our analysis shows how the spectral structure determines the temperature dependence of the QFI in different temperature windows, and identifies the spectral features that can enhance thermometric sensitivity.}
Beyond comparing specific models, our results aim to establish a general framework that links the spectral structure to fundamental bounds and optimality in equilibrium quantum thermometry, stating explicitly the conditions under which each asymptotic law holds, and to make transparent which spectral feature (degeneracy, gap, bandwidth, or growth of the density of states) controls the precision in each temperature window. Compared to previous works related to optimal equilibrium thermometry \cite{correa2015individual}, or about general theoretical laws \cite{mehboudi2019thermometry,Paris2016}, our contribution is threefold, as we here provide: (i) a unified collection of exact QFI expressions for specific physical models together with asymptotic laws (see Table \ref{tab:unified_models})\textcolor{black}{, whose caption specifies the origin of each entry}; (ii) a systematic analysis of discrete, continuous, and hybrid spectra that makes explicit when the low- or high-temperature QFI decays as $T^{-4}$ or as $T^{-2}$\textcolor{black}{; for hybrid spectra, beyond these general laws, we also derive the form of low-temperature QFI, the bound-state correction to the high-temperature law and the bound-continuum crossover temperature}; and (iii) an explicit realization of the optimal spectrum of Ref.~\cite{correa2015individual} through a fully connected quantum walk \cite{candeloro2021role} \textcolor{black}{ and, for the ring topology, an analysis of the limits $T\to0$ and $N\to\infty$ that extends Refs.\cite{Paris2016,candeloro2021role}}, together with other physical implementations of suitable quantum thermometers, spanning systems with discrete, hybrid, and continuous spectra. \textcolor{black}{For equilibrium thermometry, the complete graph topology has already been shown to possess the optimal spectral structure~\cite{correa2015individual}, and its exact QFI was derived in~\cite{candeloro2021role}. We therefore regard this case as an application of these established results, while focusing on the additional thermometric insights provided by our analysis of different models. The comparisons throughout this work concern the temperature scaling of the QFI across different spectral models. Our aim is to identify how spectral structures influence thermometric sensitivity in different temperature regimes, and all comparisons are framed in terms of asymptotic QFI scaling.}

The paper is organized as follows. In Section \ref{sec:TS}, we briefly review the description of thermal quantum states, with a focus on general limits of quantum Gibbs states and on the coupling between a temperature probe and its environment. In Section \ref{sec:FaQFI}, we examine the main theoretical tools of quantum thermometry, introducing the concepts of classical and quantum Fisher information, which intrinsically define the precision achievable in a temperature estimation protocol, and highlighting their relation to the thermometer's energy spectrum. In Section \ref{sec:R}, we compare different microscopic systems as resources for temperature estimation and analyze how spectral properties influence the scaling of precision. In Section \ref{sec:S}, we summarize the results, state the validity regime of each asymptotic law, and compare the probes under fixed-resource constraints (bandwidth or dimension). Section \ref{sec:C} closes the paper with some concluding remarks. Additional details on the derivation of the general form of the QFI, as well as its high- and low-temperature limits, for specific models, can be found in Appendix \ref{a:AMD}.

\section{Thermal States}
\label{sec:TS}
Consider an $N$-dimensional quantum system with Hamiltonian
\begin{equation}
    \label{eq:hamiltonian}
    H = \sum_{n=0}^{N-1} E_{n} \ket{\phi_{n}}\bra{\phi_{n}}.
\end{equation}
When such a system is in thermal equilibrium with an environment at temperature $T$, its state is time-independent and given by the Gibbs state
\begin{equation}
    \label{eq:thermal_rho}
    {\rho}_{th} = \frac{1}{\mathcal{Z}} \sum_{n=0}^{N-1} e^{-\beta E_{n}} \ket{\phi_{n}}\bra{\phi_{n}},
\end{equation}
where $\beta = (k_B T)^{-1}$ and $\mathcal{Z} = \Tr[e^{-\beta H}]$ is the partition function. Throughout this work, we use natural units and set $ \hbar=k_B=1 $.

The Gibbs state exhibits simple limiting behavior at the extreme temperatures that depends only on the spectrum's general structure. If the ground state is non-degenerate, then as $T\to0$ the state approaches the ground state:
\begin{equation}
    \label{eq:low temperature}
    \lim_{T\rightarrow0} \rho_{th} = \ket{\phi_{0}} \bra{\phi_{0}}.
\end{equation}
Conversely, if the Hilbert space is finite-dimensional, then as $T\to\infty$ the state tends to the maximally mixed state:
\begin{equation}
    \label{eq:high temperature}
    \lim_{T\rightarrow \infty } \rho_{th} = \frac{1}{N} \sum_{n=0}^{N-1} \ket{\phi_{n}} \bra{\phi_{n}}.
\end{equation}
Between these limits, the occupation of energy levels depends on both $T$ and the system's spectral properties. For infinite spectra (see, for example, harmonic oscillators, rotors, free particles in Sec.~\ref{sec:R}), the limit of Eq.~\eqref{eq:high temperature} does not exist, and the behavior is controlled by the density of states. Independently of the nature of the spectrum, the precise determination of the temperature of the environment requires complete knowledge of the Hamiltonian of the environment $H^{e}$, the Hamiltonian of the thermometer $H^{t}$, and their interaction $V^{e-t}$. The global system is described by
\begin{equation}
    \label{eq:environment + thermometer Hamiltonian}
    H^{g} = H^{e} \otimes \mathbb{1}^{t} + \mathbb{1}^{e} \otimes H^{t} + \lambda V^{e-t},
\end{equation}
where $\lambda$ quantifies the coupling strength. The equilibrium state of the joint system is then the Gibbs state diagonal in the eigenbasis of $H^{g}$:
\begin{equation}
    \label{eq:total gibbs}
    \rho^{g}_{th} = \frac{1}{\Tr[e^{-\beta H^{g}}]} \sum_{n=0} e^{-\beta E^{g}_{n}} \ket{\phi^{g}_{n}}\bra{\phi^{g}_{n}},
\end{equation}
with the superscript $g$ denoting the eigenvalues and eigenvectors of $H^{g}$.

However, in the weak-coupling regime ($\lambda$ sufficiently small for a given $\beta$), the reduced state of the thermometer can be approximated by its local Gibbs state. Tracing out the environment yields \cite{cresser2021weak} 
\begin{equation}
    \label{eq:approximation thermal state}
    \Tr_{e}\left(\rho^{g}_{th}\right) \simeq \rho^{t}_{th} + \mathcal{O}(\lambda),
\end{equation}
where
\begin{equation}
    \label{eq:thermal state thermometer}
    \rho^{t}_{th}= \frac{1}{\mathcal{Z}} \sum_{n=0} e^{-\beta E^{t}_{n}} \ket{\phi^{t}_{n}}\bra{\phi^{t}_{n}}.
\end{equation}
Thus, under weak coupling, the temperature $T$ of the environment can be estimated using only the spectral properties of the quantum thermometer itself. The precision of temperature estimation depends on the thermometer's energy-level structure, and different thermometer models may perform optimally in different temperature regimes. Motivated by this, we analyze several physical models with potential thermometric applications, aiming to derive general results that are valid in the weak-coupling regime. Specifically, we consider quantum systems that probe the environment's temperature under the approximation in Eq.~\eqref{eq:approximation thermal state}. For simplicity, we henceforth omit the superscript $t$ and denote the thermometer's Hamiltonian and Gibbs state as $H$ and $\rho_{th}$, respectively.

\section{Fisher and quantum Fisher information}
\label{sec:FaQFI}
Temperature is usually estimated rather than measured, i.e., it is inferred from measurements of observables whose values are temperature-dependent. This is even more true in quantum mechanics, where temperature is not an observable in a strict sense. Given a preparation of a quantum system, the measurement of an observable $Y$, and an estimator $\hat T$ of temperature, the precision of the estimation strategy is bounded by the Cram\`er-Rao theorem, as 
\begin{equation}
\label{eq:CRB}
\text{Var} \, \hat T \geq  \frac{1}{M \hbox{F}_Y(T)} ,
\end{equation}
where $M$ is the number of measurements performed on the system, and
\begin{equation}
\label{eq:fisher_information}
\hbox{F}_{Y}(T) = \sum_{y} \dfrac{\left| \partial_{T} p \left( y \vert T \right) \right|^{2}}{p\left( y \vert T \right)},
\end{equation}
is the Fisher information. The probability density $p\left( y \vert T \right)$ describes the distribution of the outcomes of $Y$ given the value of temperature, and is obtained by Born's rule. The supremum of the FI, i.e., the maximization over all possible quantum measurements, is termed quantum Fisher information (QFI) and may be explicitly calculated as
\begin{equation}
    \label{eq:general_qfi}
    \mathcal{F}(\lambda) = \hbox{Tr}[\rho_{\lambda}\mathcal{L}^{2}_{\lambda}],
\end{equation}
where $\mathcal{L}$ is the so-called symmetric logarithmic derivative, defined as 
\begin{equation}
    \label{eq:sld}
    \partial_{\lambda} \rho_{\lambda} = \frac{ \mathcal{L}_{\lambda}\rho_{\lambda} + \rho_{\lambda} \mathcal{L}_{\lambda} }{2}.
\end{equation}
For the estimation of the temperature, the optimal measurement corresponds to the energy of the system, i.e., 
$\hbox{F}_{E}(T)=\mathcal{F}(T)$. \textcolor{black}{For a state at thermal equilibrium, the quantum Fisher information is given by the fluctuations of the Hamiltonian}
\begin{equation}
    \label{eq:(q)fi}
    \hbox{F}_{E}(T)=\mathcal{F}(T)= \frac{\langle {(\Delta H)^{2}} \rangle}{T^{4}} = \frac{C_V(T)}{T^2}.
\end{equation}
\color{black}
For a system in thermal equilibrium, the temperature is encoded in its Gibbs state \eqref{eq:thermal state thermometer}. Given that the Hamiltonian is temperature-independent, the symmetric logarithmic derivative is given by $\mathcal{L}_T=(H-\langle H\rangle_T)/T^2$. \textcolor{black}{Indeed, since $[H,\rho_{\rm th}]=0$, one has $\partial_T\rho_{\rm th}=\rho_{\rm th}\,(H-\langle H\rangle_T)/T^2$, so that this $\mathcal{L}_T$ solves Eq. \eqref{eq:sld} and, inserted in Eq. \eqref{eq:general_qfi}, yields Eq. \eqref{eq:(q)fi}.} It is therefore diagonal in the energy eigenbasis, thus, an energy measurement is optimal for estimating the temperature. As a result, the classical Fisher information obtained from an energy measurement equals the QFI. Here, $\langle(\Delta H)^2\rangle=\langle H^2\rangle-\langle H\rangle^2$ denotes the energy variance evaluated in the thermal equilibrium state, while $C_V(T)=\partial_T\langle H\rangle$ is the heat capacity. The last equality in Eq.~\eqref{eq:(q)fi} follows from 
\begin{equation}
C_V(T)=\frac{\partial\langle H\rangle}{\partial T} = \frac{\langle H^2\rangle-\langle H\rangle^2}{T^2}.
\end{equation}
At thermal equilibrium, the temperature-estimation precision is directly determined by the energy fluctuations of the system, which are proportional to its heat capacity. A larger heat capacity implies a higher thermometric precision.
\color{black}
Even if Eq.~\eqref{eq:(q)fi} depends on the Hamiltonian, for a discrete non-degenerate spectrum, we have general expressions valid in low- and high-temperature limits. In the low-temperature limit, i.e., a temperature range for which the only occupied energy levels are the ground state $\ket{\phi_{0}}$ and the first excited state $\ket{\phi_{1}}$ (i.e., $e^{-\beta E_{2}} \simeq 0$), the temperature QFI may be written as
\begin{eqnarray}
    \label{eq:limit qfi small T}
     {\cal F}(T) \stackrel{T \ll E_{2}}{\simeq} \frac{\left( E_{1} - E_{0} \right)^2}{4 T^4 
     \cosh^2\left[ (E_{1} - E_{0})/2T\right]}\,.
\end{eqnarray}
In the high-temperature limit, and still considering a system with a finite spectrum (i.e. $e^{-\beta E_{N-1}} \simeq 1$), we may write
\begin{align}
    \label{eq:limit qfi infinite T}
 {\cal F}(T) \stackrel{T \gg E_{N-1}}{\simeq} & \frac{\left[ \frac{1}{N} \sum_{n} E_{n}^{2} - \frac{1}{N^2} \left( \sum_{n} E_{n} \right)^2 \right]}{T^{4}} \,.
\end{align}
Indeed, in the large-$T$ behavior, the QFI of finite spectra is fixed by $\mathrm{Var}(H)$ or, equivalently, by $C_V$. Taking into account continuous or degenerate spectra, the behavior of the QFI and its low- and high-temperature limits may vary considerably. Indeed, a systematic study of the spectral properties of quantum probes is fundamental to the design of precise quantum thermometers.


\section{Results}
\label{sec:R}
The results of Sec.~\ref{sec:FaQFI} indicate that the spectral distribution of energy levels in a quantum system determines how precisely the environmental temperature can be estimated, with different quantum systems yielding different precision. As an example, confined and free quantum particles exhibit different spectra, ranging from discrete to continuous, but temperature information is encoded in the thermal occupation of energy levels and is determined by the density of states. This means that, independently of the physical realization, the thermometric performance is mainly governed by the spectral structure of the quantum system itself.
\begin{figure}[!htb]
    \centering
    \includegraphics[width=1\linewidth]{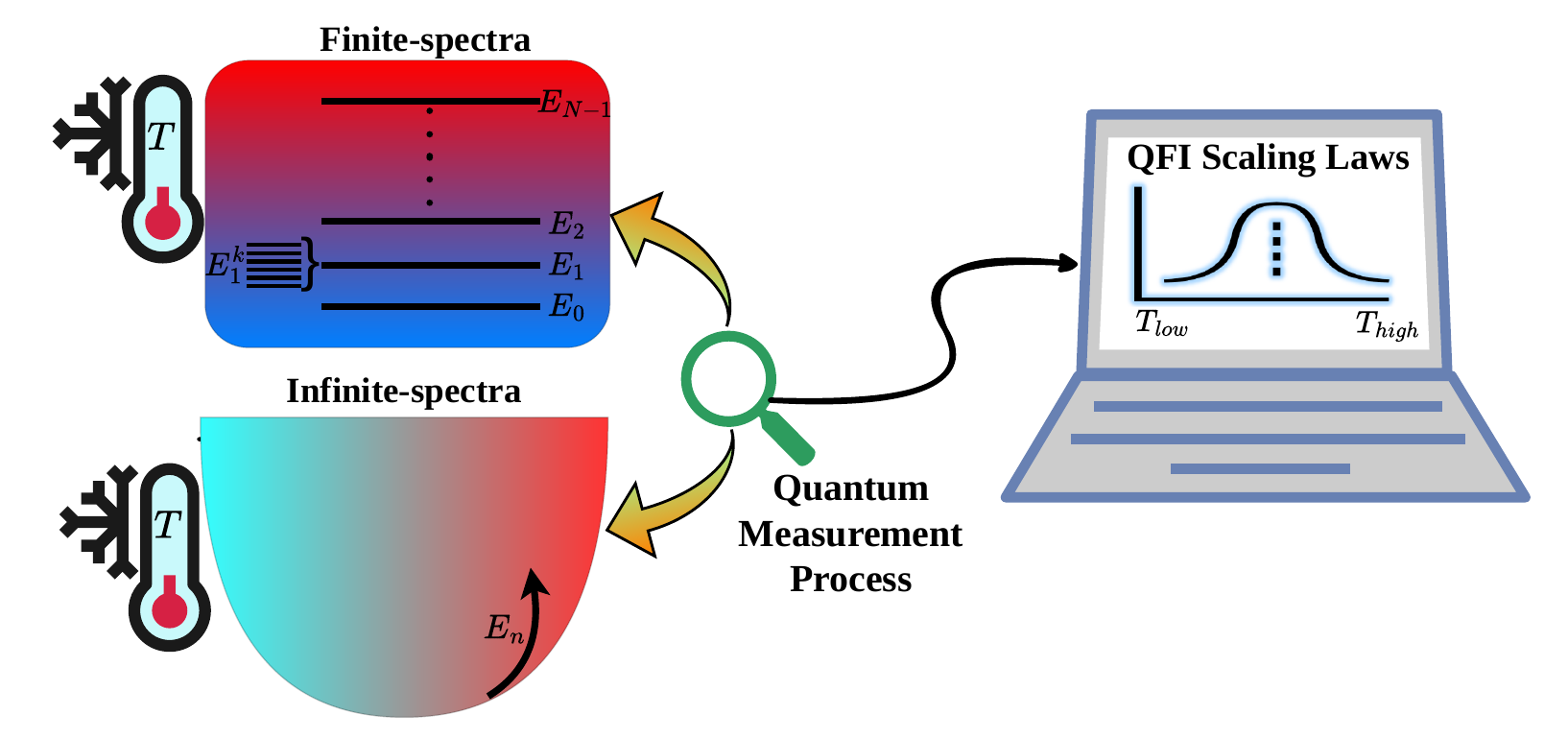}
    \caption{Schematic of quantum thermometry for systems with finite and infinite/continuous energy spectra. A thermal environment at temperature T populates the probe’s energy levels, and temperature information is extracted through measurements quantified by the quantum Fisher information.}
    \label{fig:QT_Schematic}
\end{figure}
Indeed, in the following sections, we will analyze spectra of paradigmatic physical systems and investigate how the energy-level spacing affects the (Q)FI and its scaling across various temperature regimes. Throughout, scaling laws denote the leading asymptotic behavior of the QFI in a specified limit; exact expressions are explicitly identified, while asymptotic results and approximations are stated together with the limits and regimes in which they are valid. After the discussion of each specific model, in Sec.~\ref{sec:S} we provide a meaningful comparison between different thermometers, specifying the energy scale (gap, bandwidth) and the resource considered for each probe (number of levels, degeneracy, coupling).

\subsection{Quantum thermometry using probes with finite spectrum}
Let us first consider paradigmatic quantum systems that have a finite spectrum.
\subsubsection{Spin Systems} We begin this analysis with $N$ dimensional spin systems. Specifically, for a spin-$s$ particle in a uniform magnetic field $B$ along the \(z\)-axis, the Hamiltonian can be written as
\begin{equation}\label{key}
	H^{s} = -\omega S_z, \quad \omega = \gamma B > 0,
\end{equation}
and the associated energy spectrum is
\begin{equation}
    \label{eq:spin energy spectrum}
	E^{s}_m = -\omega m, \quad m = -s, -s+1, \dots, s \, ,
\end{equation}
i.e. $N = 2s+1$ equally spaced levels with spacing $\Delta = \omega$. Each energy level corresponds to a Zeeman sub-level of the spin projection along $B$, the spectrum is symmetric and evenly spaced.
The energy variance $\mathrm{Var}(H^{s})$ can be calculated analytically, and the QFI for the temperature is given by
\begin{eqnarray}
    \mathcal{F}^{s} (T)&=& \frac{\omega^2}{4 T^4} \left[ \mathrm{csch}^2\left( \frac{\omega}{2 T} \right) - N^2 \mathrm{csch}^2\left( \frac{N\omega}{2 T} \right) \right].\nonumber\\
\end{eqnarray}
This QFI expression is valid for any probe with $N$ equally spaced levels, upon replacing $\omega$ with the level spacing $\Delta$. The expression highlights the role of spectral boundedness. In turn, the first term has the same functional form as the QFI of a single quantum harmonic oscillator, see Eq.~\eqref{QFI_QHO}, reflecting the equally spaced nature of the energy levels. The second term, which depends explicitly on $N$, originates from the finite-dimensional (boundedness) nature of the spectrum. In fact, finite spin systems can be interpreted as truncated harmonic oscillators that continuously approach the behavior of the harmonic oscillator in the limit $N \to \infty$. Let us now address the asymptotic behavior of the QFI at low and high temperature (see App.~\ref{SS:AMD}). At low temperatures, the QFI behaves as
\begin{equation}
    \label{eq:low T spin}
	\mathcal F^{s}(T)
	\simeq
	\frac{\omega^2}{T^4}
	e^{-\omega/T},
\end{equation}
which depends only on the lowest energy gap (see Eq.~\eqref{eq:limit qfi small T}).
For $T \to \infty$ we have
\begin{equation}
	\mathcal F^{s}(T)
	\simeq
	\frac{\omega^2}{12 T^4}
	\left(N^2 - 1\right).
\end{equation}
At low temperature, the QFI is exponentially suppressed because only the first excited level contributes, whereas for $T\gg\omega N$ it decays as $T^{-4}$, with a prefactor proportional to $\omega^2 N^2$. Following Eq.~\eqref{eq:spin energy spectrum} and Eq.~\eqref{eq:limit qfi infinite T}, this $N^2$ factor reflects the bandwidth growth rather than an intrinsic benefit of the dimension.

\subsubsection{$N$-level atom with ground state + $(N-1)$ degenerate excited states} 
Here, we consider an atom with a non-degenerate ground state and an $ (N -1) $-fold degenerate excited level Hamiltonian, 
\begin{equation}
	H^{N} = E_0 \ket{\phi_{0}}\bra{\phi_{0}} + E_{1} \sum_{n=1}^{N-1} \ket{\phi_{n}}\bra{\phi_{n}} \, .
\end{equation}
In this way, we can examine how energy-level degeneracy affects the precision of temperature estimation in discrete spectra. This spectrum was shown in \cite{correa2015individual} to be the optimal one among all $N$-level Hamiltonians for equilibrium thermometry at a given temperature, once the gap is suitably tuned. Here we recall its properties, which are used below to characterize the role of degeneracy in discrete spectra.
Without loss of generality, the energy spectrum can be written as
\begin{equation}
   \label{eq:spectrum_atom_degenerate}
	E^{N}_0 = 0, \quad E^{N}_1 = E^{N}_2 = \cdots = E^{N}_{N-1} = \epsilon.
\end{equation}
The energy variance of such a model is
\begin{equation}
	\mathrm{Var}(H^N) = \frac{(N-1)\epsilon^2 e^{-\beta\epsilon}}{[1 + (N-1)e^{-\beta\epsilon}]^2},
\end{equation}
and consequently the QFI for temperature,
\begin{equation}
    \label{eq:fisher_atom_degenerate}
	\mathcal{F}^{N}(T) = \frac{1}{T^4} \cdot \frac{(N-1)\epsilon^2 e^{-\beta\epsilon}}{[1 + (N-1)e^{-\beta\epsilon}]^2}.
\end{equation}
In the low-temperature limit, where $q\equiv(N-1)e^{-\epsilon/T}\ll1$, Equation \eqref{eq:fisher_atom_degenerate} reads $\mathcal{F}^N=\epsilon^2q/[T^4(1+q)^2]\le\epsilon^2/(4T^4)$ and, keeping the leading exponential contribution, reduces to
\begin{equation}
	\label{eq:low_T_QFI_atom_degenerate}
	\mathcal F^{N}(T)
	\simeq
	\frac{(N-1)\epsilon^2}{T^4}
	e^{-\epsilon/T}.
\end{equation}
Thus, for $q\ll1$ the sensitivity is exponentially suppressed by the spectral gap $\epsilon$, while the prefactor is enhanced by the excited-state degeneracy $(N-1)$ (compare with Eq.~\eqref{eq:low T spin} to explicitly see the effect of degeneracy). The enhancement is bounded: at fixed $T$ and $\epsilon$ the QFI cannot exceed $\epsilon^2/(4T^4)$ and, for $q\gg1$, it decreases as $1/(N-1)$.

For high temperature $T \to \infty$ ($\beta \to 0$), the Gibbs state approaches a uniform distribution over the $N$ levels. Expanding the Boltzmann factor to the leading order, $e^{-\beta\epsilon}\simeq 1 + \mathcal{O}(\beta)$, the partition function becomes $ Z \simeq N $, and the QFI simplifies to
\begin{equation}
	\label{eq:high_T_QFI_atom_degenerate}
	\mathcal F^{N}(T)
	\simeq
	\frac{(N-1)\epsilon^2}
	{N^2 T^4}.
\end{equation}
At high temperatures, the QFI decays as $T^{-4}$, and the $N$-dependent prefactor shows that strong degeneracy does not lead to quadratic enhancement at high temperatures. Overall, for $q\ll1$ the QFI grows linearly with $N$, whereas for $T\gg\epsilon$ it decreases as $N^{-1}$.
This trade-off implies that, for fixed temperature and gap, there exists an optimal number of levels. Indeed, maximizing Eq.~\eqref{eq:fisher_atom_degenerate} with respect to $N$ gives
\begin{equation}
    \label{eq:N_opt atom}
    N_{\mathrm{opt}}(T/\epsilon) = 1 + e^{\epsilon/T},
\end{equation}
at which the QFI reaches its maximum,
\begin{equation}
    \label{eq:QFI_opt atom}
    \mathcal{F}_{\mathrm{max}}(T) = \frac{\epsilon^2}{4T^4}.
\end{equation}
Below this optimal size, adding more levels increases the sensitivity. Above it, adding more levels makes things worse. It is worth noting how this optimal size changes with temperature. At low temperatures, the exponential $e^{\epsilon/T}$ becomes extremely large, so $N_{\mathrm{opt}}$ is enormous, i.e., many degenerate states are needed. As the temperature increases, $N_{\mathrm{opt}}$ shrinks with the exponential approaching one, and $N_{\mathrm{opt}}$ tends to 2, meaning that a simple two-level system is already optimal. Degeneracy is thus a resource, but a bounded one; its usefulness depends on the temperature window, and it cannot be increased indefinitely.

\subsubsection{Quantum walks} When a quantum system can be modeled as a collection of $N$ qubits (nodes) that form a network connected by a specific topology, it may be described using the quantum walk (QW) formalism. 
The system is described by an excitation-preserving Hamiltonian in the form
\begin{align}
    \label{eq:total_Hamiltonian_QWs}
    H^{QW}=\sum_{i=0}^{N-1}\omega_i\,\sigma_i^+\sigma_i^-+\sum_{i<j}g_{ij}\left(\sigma_i^+\sigma_j^-+\sigma_j^+\sigma_i^-\right),
\end{align}
where $\sigma_i^\pm$ are the ladder operators of qubit $i$, $\omega_i$ is the on-site energy and $g_{ij}$ the hopping strength between nodes $i$ and $j$, each pair being counted once. In the single-excitation regime and under the condition of isotropy, $g_{ij}=-\mathcal{J}$ for connected nodes ($g_{ij}=0$ otherwise) and $\omega_i=\omega\,\deg(i)$, i.e., proportional to the degree (number of connections) of node $i$, the QW Hamiltonian reduces to
\begin{align}
    \label{eq:Hamiltonian_QWs_se}
    \bra{i}{H}^{QW}\ket{j} = \begin{cases}
    \omega\,\deg(i) & \text{if } i=j,\\
    -\mathcal{J} & \text{if } i\neq j \text{ are connected},\\
    0 & \text{otherwise.}
    \end{cases}
\end{align}
Under this model, we assume that bath exchanges energy, but not excitations, with the QW, i.e., that the system-bath interaction commutes with the excitation-number operator $\sum_i\sigma_i^+\sigma_i^-$. Within this assumption, the Gibbs state of the model is then entirely described by Eq.~\eqref{eq:Hamiltonian_QWs_se}, through the standard thermal relation $\rho_{\rm th}=e^{-\beta H^{QW}}/\mathcal{Z}$. As for all the other models also in this condition, the optimal measurement consists of the measurement of the eigenbasis of ${H}^{QW}$. Then, different topologies are associated with different Hamiltonians and consequently with different thermometric properties. A topology in which all the qubits are connected to each other (complete graph, $\deg(i)=N-1$) has the Hamiltonian
\begin{equation}
    \label{eq:H_Complete_Network}
    H^{c} = (N-1) \omega \mathbb{1} - \mathcal{J}\sum_{i>j} \ket{i}\bra{j} + \ket{j}\bra{i},
\end{equation}
and was proven to be a useful resource in previous estimation applications of the model \cite{candeloro2020continuous} \textcolor{black}{, and its thermometric QFI was derived in Ref.\cite{candeloro2021role}}. Its spectrum consists of a non-degenerate ground state, the uniform superposition of the sites, with energy $E_0=(N-1)(\omega-\mathcal{J})$, and of an $(N-1)$-fold degenerate excited level with energy $E_1=(N-1)\omega+\mathcal{J}$, so that the gap is $\Delta=E_1-E_0=N\mathcal{J}$. The spectrum (up to an irrelevant energy shift) is then exactly the same as Eq.~\eqref{eq:spectrum_atom_degenerate}, with $\epsilon=N \mathcal{J}$. The thermometric properties of the complete graph are therefore those of the $N$-level atom, i.e., Eq.~\eqref{eq:fisher_atom_degenerate}, with the limits provided in Eq.~\eqref{eq:low_T_QFI_atom_degenerate} and Eq.~\eqref{eq:high_T_QFI_atom_degenerate}, together with the optimum provided in Eq.~\eqref{eq:N_opt atom},~\eqref{eq:QFI_opt atom}. The complete graph thus provides a physical realization of the optimal equilibrium thermometer of Ref.~\cite{correa2015individual}. Since $\Delta = N\mathcal{J}$, the gap and degeneracy cannot be varied independently in the fully connected topology. Increasing $N$ at fixed $\mathcal{J}$ enhances the excited state degeneracy but simultaneously increases the spectral gap. To separate the effect of degeneracy from that of the gap, comparisons should be made at fixed $\Delta$, corresponding to $\mathcal{J}\propto N^{-1}$, for which the low-temperature QFI scales linearly with $g_1=N-1$ (for $(N-1)e^{-\Delta/T}\ll1$). This means that under the conditions expressed in \cite{correa2015individual}, a QW model with a fully connected topology can be adopted as the optimal probe for quantum thermometry, but the coupling must therefore be re-tuned when the temperature changes.\par

The thermometric performance of the ring (cycle) graphs, together with that of the complete graphs, was analyzed in \cite{candeloro2021role}. Here we recall their spectrum and focus on the size dependence of the gap, together with the explicit derivation of analytical limits $T\to\infty$ and $T\to0$. The Hamiltonian of the ring topology can be expressed as
\begin{equation}
    \label{eq:H_Ring}
    H^{r} = 2 \omega \mathbb{1} - \mathcal{J}\sum_{i=0}^{N-1} \ket{i}\bra{i+1} + \ket{i+1}\bra{i},
\end{equation}
with the periodic boundary conditions $\ket{N}=\ket{0}$. The energy levels are
\begin{equation}
    {E}^{r}_{l} = 2 \omega -2 \mathcal{J}\cos \left( \frac{2\pi(l+1)}{N} \right) \; ; \; l = 0,...,N-1 .
\end{equation}
Here we will also focus on the study of the role of $\mathcal{J}$, as it was proven to be a fundamental resource in other physical applications \cite{coutinho2025peak,cavazzoni2025perfect}. By changing the value of $\mathcal{J}$, it is possible to modify the gap between two successive energy levels and to enlarge or reduce the bandwidth of the system, i.e., the difference between the maximum and the minimum eigenvalue of the system’s Hamiltonian. The temperature QFI of such an energy spectrum is
\begin{align}
    \label{eq:variance ring qw}
    &\mathcal{F}^{r}(T) = \frac{4\mathcal{J}^2}{T^4} \bigg[ \frac{\sum_{l=0}^{N-1} \cos^2\left(\frac{2\pi(l+1)}{N}\right) e^{2\beta\mathcal{J} \cos\left(\frac{2\pi(l+1)}{N}\right)}}{\sum_{l=0}^{N-1} e^{2\beta\mathcal{J} \cos\left(\frac{2\pi(l+1)}{N}\right)}} \nonumber \\
    &- \left( \frac{\sum_{l=0}^{N-1} \cos\left(\frac{2\pi(l+1)}{N}\right) e^{2\beta\mathcal{J} \cos\left(\frac{2\pi(l+1)}{N}\right)}}{\sum_{l=0}^{N-1} e^{2\beta\mathcal{J} \cos\left(\frac{2\pi(l+1)}{N}\right)}} \right)^2 \bigg]\,.
\end{align}
The low- and high-temperature limits of Eq.~\eqref{eq:variance ring qw} capture the essential thermometric behavior of the model. In the low temperature regime $T \to 0$, only the ground and first excited levels contribute. \color{black}The latter is two-fold degenerate, and the energy gap scales as $\Delta = 4\mathcal J\sin^2\left(\frac{\pi}{N}\right)$, with a two-fold degeneracy of the first excited state, the QFI is $\mathcal{F}^{r}(T)=2\Delta ^{2}e^{-\Delta/T}/T^{4}$~\cite{candeloro2021role}. Expanding $\Delta \simeq 4\pi^{2}\mathcal{J}/N^{2}$ for large $N$ gives \color{black}
\begin{equation}
	\mathcal{F}^{r}(T) \simeq \frac{32\pi^4 \mathcal{J}^2}{N^4 T^4}
	\exp\!\left[-\frac{4\pi^2 \mathcal{J}}{N^2 T}\right].
\end{equation}
This expression shows a competition between two mechanisms that depend on $N$: increasing $N$ reduces the gap and improves thermal accessibility, whereas the prefactor scales as $N^{-4}$. As a consequence, for fixed temperature $T$, the QFI exhibits a non-monotonic dependence on $N$ with an optimal system size determined by the balance between exponential suppression and algebraic suppression.\par
In the opposite high-temperature regime $T \to \infty$ at fixed $N$, all levels become equally populated, and the QFI reduces to a finite-size expression independent of the spectral resolution
\begin{equation}
	\mathcal{F}^{r}(T) \simeq \frac{2\mathcal{J}^2}{T^4} \, ,
\end{equation}
with this limit that holds for $N\ge3$ and $T\gg J$. The interesting aspect of this model is that by varying $N$, one may observe the transition from a discrete to a continuous spectrum.
Indeed, if the thermodynamic limit $N \to \infty$ is taken before the low and high temperature expansions, Eq.~\eqref{eq:variance ring qw} rewrites as
\begin{equation}
    \label{eq:variance ring qw th limit}
    \mathcal{F}^{r}(T) \simeq \frac{4\mathcal{J}^2}{T^4} \left[ \frac{1}{2} \left( 1 + \frac{I_2(2\beta\mathcal{J})}{I_0(2\beta\mathcal{J})} \right) - \left( \frac{I_1(2\beta\mathcal{J})}{I_0(2\beta\mathcal{J})} \right)^2 \right],
\end{equation}
where $I_n(x)$ denotes the modified Bessel function of the first kind. The low- and high-temperature limits of the above expression are, respectively,
\begin{equation}
    \label{eq:low T ring}
    \mathcal{F}^{r}(T) \simeq \frac{1}{2T^2} \, ,
\end{equation}
\textcolor{black}{These limits differ, taking the limit $N\to\infty$ first lets the higher levels contribute and yields the $1/(2T^2)$ of Eq.~\eqref{eq:low T ring}, whereas the finite-$N$ result holds when the first excited level dominates, thus its large-$N$ form is not the full continuum limit. And the high-T is}
\begin{equation}
    \label{eq:high T ring}
    \mathcal{F}^{r}(T) \simeq \frac{2\mathcal{J}^2}{T^4} \, .
\end{equation}
These low- and high-temperature behaviors refer to Eq. ~\eqref{eq:variance ring qw th limit}, and the limits are expressed in terms of $\mathcal{J}$, i.e., they have to be interpreted as $T\ll\mathcal{J}$ and $T\gg\mathcal{J}$. In the thermodynamic limit, the spectrum becomes continuous but remains bounded. Consequently, the QFI retains the $T^{-4}$ high-temperature scaling. This implies that a continuous spectrum alone does not ensure a scaling of $T^{-2}$. \textcolor{black}{The role of a tunable energy gap in thermometry was discussed in~\cite{Paris2016}, and its realization in the cycle graph was studied in~\cite{candeloro2021role}. The present results examine how tuning $\mathcal{J}$ and varying the system size influence the QFI in both the finite-ring and thermodynamic limits, and show different scaling behaviors that arise when these limits are taken in different orders.}

All finite-bounded spectrum probes considered in this work exhibit the same qualitative thermometric behavior. At low temperatures, in agreement with the general limits derived in Refs.~\cite{Paris2016,CorreaPhysRevB2018,potts2019fundamental,AiachePRAlowT}, the QFI is exponentially suppressed and controlled by the smallest gap and by the degeneracy of the first excited level. In this context, quantum walks on fully connected networks provide a way to increase low-temperature QFI via the degeneracy of the first excited state, while the exponential dependence on the gap remains unchanged. On the other hand, for quantum walks on ring topologies, by tuning the system size, the first energy gap closes as $N^{-2}$ and can be tuned through the size, 
leading at low temperatures to a possible power-law scaling $T^{-2}$ instead of the standard exponential suppression. At high temperatures, the QFI decays algebraically and is determined by the energy variance 
\begin{equation}\label{eq:Var_bound}
    \mathrm{Var}(H)\le \frac{(E_{\max}-E_{\min})^2}{4} \, ,
\end{equation}
in any state. Hence, the sensitivity is ultimately limited by the probe's spectral bandwidth. Therefore, any enhancement with increasing system size at fixed level spacing arises only from the increase in the bandwidth. The bound is saturated, at a given temperature, by the spectrum of Eq.~\eqref{eq:spectrum_atom_degenerate} with $N-1=e^{(E_{\max}-E_{\min})/T}$.

\subsection{Quantum thermometry using confined probes} 

As a second prototypical class of models, we focus on systems defined by a confining potential, such as quantum wells and harmonic oscillators. These models are well-suited to describe a wide variety of physical systems and quantum devices \cite{forghieri2023quantum,fanucchi2025giant}. Additionally, they provide examples of systems with an infinite spectrum or, under certain approximations, systems with an unbounded continuous spectrum (see the following subsections for further details).

\subsubsection{Quantum harmonic oscillator} As a first confining potential, we look at the quantum harmonic oscillator (QHO), as it represents a ubiquitous model in quantum sensing and generally in physics \cite{leitch2022driven,cavazzoni2025frequency,longhi2025mpemba}. The Hamiltonian of a harmonic oscillator $k$ can be written in the form 
\begin{equation}
    \label{eq:QHO}
    H^{qho}_{k} = \frac{p_k^2}{2m} + \frac{1}{2} m\omega^2 x_k^2,
\end{equation}
describe, at least to a first approximation, any vibrational modes of a quantum system. For 
$N$ independent identical oscillators of frequency $\omega$, the Hamiltonian can be written as
\begin{equation}
    H_{QHO}=\sum_{k=1}^{N}H^{qho}_k=\sum_{k=1}^{N}\sum_{n=0}^{\infty}\omega\left(n+\tfrac12\right)|n\rangle_k\langle n|,
\end{equation}
and the energy variance is additive:
\begin{equation}
\label{eq:additive}
	\mathrm{Var}(H^{QHO}) = \sum_{k=1}^{N} \mathrm{Var}(H^{qho}_k)
	= N \omega^2 \bar{n}(\bar{n}+1).
\end{equation}
where $\bar n=\langle\hat n_k\rangle=(e^{\omega/T}-1)^{-1}$ is the mean thermal occupation of each oscillator. The total QFI is then given by
\begin{equation}\label{QFI_QHO}
	\mathcal{F}^{QHO}_{\omega}(T) = N \frac{\omega^2}{4 T^4} \, \mathrm{csch}^2\left( \frac{\omega}{2 T} \right).
\end{equation}
For a multi-mode system with different frequencies $\omega_k$, we have  
\begin{equation}
	\mathcal{F}^{QHO}_{\omega_k}(T)=\frac{1}{T^4}\sum_{k=1}^N
   \omega_k^2  \frac{e^{\beta\omega_k}}{\big(e^{\beta\omega_k}-1\big)^2}.
\end{equation}
If we assume sufficiently large modes, we can apply the thermodynamic limit. For a continuous set of modes with frequencies $\omega_k$ and density of states $g(\omega)$,
replacing the sum with an integral, the QFI reads
\begin{equation}
	\mathcal{F}^{QHO}_{\text{cont}}(T) = \frac{1}{T^4} \int_0^{\omega_c} g(\omega) \omega^2 	\frac{e^{\beta \omega}}{(e^{\beta \omega}-1)^2} \, d\omega.
\end{equation}
For a general power-law behavior of the density of states,
\begin{equation}
	g(\omega) = A \, \omega^s, \qquad 0 \le \omega \le \omega_c, \qquad s > -1
\end{equation}
with a cutoff $\omega_c$, and $A$ is a normalization constant; the variance of the Hamiltonian reads
\begin{eqnarray}
	\mathrm{Var}(H) &=& \int_0^{\omega_c} A \, \omega^s\, \omega^2 \frac{e^{\beta \omega}}{(e^{\beta \omega}-1)^2} \, d\omega.
\end{eqnarray}
Changing variable $y = \beta \omega$, $d\omega = dy/\beta$, the QFI becomes
\begin{equation}
	\mathcal{F}_{\text{cont}}(T) = A\, T^{s-1} \int_0^{\beta \omega_c} y^{s+2} \frac{e^y}{(e^y-1)^2} \, dy.
\end{equation}
At low temperatures ($ T \ll \omega_c$), the QFI scales as $\mathcal{F}_{\text{cont}}(T) \propto T^{\,s-1}$, while at high temperatures ($T \gg \omega_c$) it approaches the classical scaling of free particles, i.e., $\mathcal{F}_{\text{cont}}(T) \propto 1/T^2$ (see Sec.~\ref{subsec:Qtwfp}).
The quantum harmonic oscillator, therefore, provides a unified framework connecting discrete and continuous thermometry through the structure of its mode distribution. For a finite number of modes, the additivity of the independent oscillators produces a linear enhancement $ \mathcal{F} \sim N/T^2 $ at high temperatures, showing the characteristic scaling $1/T^2$. In the continuum limit, the QFI is governed by the density of states $g(\omega) \propto \omega^s$, at low temperatures $T \ll \omega_c$, thermal occupation is effectively restricted to low-frequency modes, so the QFI inherits the spectral weight near $\omega = 0$, giving $ \mathcal{F} \propto T^{s-1} $. At high temperatures $T \gg \omega_c$, the spectral structure becomes ``irrelevant", and the system shows a universal scaling in the form $ \mathcal{F} \propto 1/T^2 $.

\subsubsection{Quantum Rotor} 
After the description of the thermometric properties of a quantum harmonic oscillator, which can be adopted to model the vibrational modes of any quantum system, we now focus on a model that can describe rotational modes \cite{leitch2024thermodynamics}. Together, these two models represent the basis for the quantum description of the degrees of freedom of any quantum rigid system. Due to the additive property of the energy variance (Eq.~\eqref{eq:additive}), we can focus on the properties of the partition function of a single quantum rotor to study the main properties of the model. The Hamiltonian of such systems can be written as follows
\begin{equation}
	\label{eq:QR}
	H^{qr} = \frac{\hat{L}^2}{2I}  \, ,
\end{equation}
where $\hat{L}$ and $I$ are, respectively, the angular and inertia momentum. This Hamiltonian leads to energy levels in the form of
\begin{equation}
	\label{eq:eQR}
	E^{qr}_{n} = \frac{1}{2I} n(n+1) = B n(n+1)
\end{equation}
each with degeneracy $g_n = 2n +1$. The temperature QFI of a quantum rotor is
\begin{align}
	\label{eq:QFI rotor full}
	\mathcal{F}^{qr}(T) = \left( \frac{1}{2I T^2} \right)^2 &\bigg[ \frac{\sum_{n} g_n h_n^2 e^{-\beta E^{qr}_{n} } }{\sum_{n} g_n e^{-\beta E^{qr}_{n}} } - \nonumber \\
	&\left( \frac{\sum_{n} g_n h_n e^{-\beta E^{qr}_{n} } }{\sum_{n} g_n e^{-\beta E^{qr}_{n}} } \right)^2 \bigg],
\end{align}
with $h_n = n(n + 1)$.

For  $T \to 0$,  only the lowest energy levels are thermally populated. The ground state $n=0$ has energy $E_0 = 0$ and degeneracy $g_0 = 1$. The QFI in this regime thus  reads
\begin{equation}
    \mathcal{F}^{qr}(T) \simeq \frac{3\Delta^2}{T^4} e^{-\Delta/T},
\end{equation}
with degeneracy contributing an extra prefactor $g_1=3$.
For high temperature $T\rightarrow\infty$, many levels contribute, and the sum can be approximated by an integral. However, the scaling is unaffected, and at high T the QFI one has
\begin{equation}
	\mathcal{F}^{qr}(T) \simeq \frac{1}{T^2}.
\end{equation}
The quantum rotor has an infinite and unbounded spectrum, so that the number of thermally accessible states increases with temperature. As a consequence, the energy variance grows as $T^2$ and the QFI decays as $T^{-2}$ (See App.~\ref{QR:AMD} for details on the derivation of all the results related to the quantum rotor). 

\subsubsection{Quantum Wells} 
One of the easiest yet widely adopted confined models consists of considering a system with a Hamiltonian in the form
\begin{equation}
    \label{eq:QWs}
    H^{qw} = \frac{\hat{p}^2}{2m} + \hat{V}({x}_{k}),
\end{equation}
where
\begin{equation}
    \label{eq:VQWs}
    \hat{V}(x_k)=\begin{cases}0,&0<x_k<d_k,\\ \infty,&\text{otherwise.}\end{cases}
\end{equation}
Despite the simplicity of the model, this Hamiltonian provides a convenient description for semiconductor hetero-structures or quantum dots \cite{fanucchi2025giant}. Following the results presented in Eq.~\eqref{eq:additive}, we consider a $1D$ infinite quantum well of width $d$, whose energy spectrum is
\begin{equation}
	E^{qw}_n = \frac{n^2 \pi^2}{2m d^2}, \quad n = 1, 2, 3, \dots,
\end{equation}
we can simplify it by writing it as $E_n = \alpha n^2$, with $ \alpha = \frac{\pi^2}{2m d^2} $. This spectrum has a level spacing that grows linearly with $ n $, i.e. $ \Delta E_n = E_{n+1} - E_n = \alpha (2n + 1) $. The QFI for the quantum Wells model reads as
\begin{equation}
	\mathcal{F}^{qw}(T) =\frac{1}{T^4}\bigg[ \frac{ \vartheta_3''(0,e^{-\alpha \beta}) }{\vartheta_3(0,e^{-\alpha \beta}) - 1}
	- \left( \frac{ \vartheta_3'(0,e^{-\alpha \beta}) }{\vartheta_3(0,e^{-\alpha \beta}) - 1} \right)^2 \bigg],
\end{equation}
with $ \vartheta_3'(0,e^{-\alpha \beta}) $ and $ \vartheta_3''(0,e^{-\alpha \beta}) $ denoting, respectively, the first and second derivatives of the Jacobi theta function with respect to the inverse temperature $ \beta $ (see App.~\ref{QW:AMD}).

In the low temperature limit $T \to 0$, the QFI reads 
\begin{equation}
	\mathcal{F}^{qw}(T) \simeq \frac{9\alpha^2}{T^4} \, e^{-3\alpha/T}.
\end{equation}
The energy gap between the ground state and the first excited state is $E_2 - E_1 = 3\alpha$. The low-temperature scaling follows the universal form.

For the high temperature limit $T \to \infty$, i.e., $\beta \to 0$, the QFI reads
\begin{equation}
	\mathcal{F}^{qw}(T) \simeq \frac{1}{2T^2}.
\end{equation}
Thus, the thermometric sensitivity exhibits the universal high-temperature scaling characteristic of systems with effectively unbounded spectra.

\subsubsection{Power-law spectral model}
After characterizing the thermometric performance of different confined probes, a key question is how the QFI of such systems scales with system size and the structure of the energy spectrum. In confined thermometry, where quantum states are trapped in finite domains, the energy spectrum is inherently discrete, and its growth with the quantum number $n$ determines the ultimate temperature sensitivity. To systematically investigate this, we consider a family of models in which the energy levels follow a power law
\begin{equation}
	E_n = C n^\alpha, \qquad n = 0,1,2,\dots, N-1,
\end{equation}
with $C > 0$ setting the overall energy scale, $\alpha > 0$ determining the growth rate of the spectrum, and $N$ the total number of bound states (which may be finite or infinite). 

This spectral form naturally emerges in confined systems: $\alpha=1$ corresponds to evenly spaced levels such as those of a harmonic oscillator, while $\alpha=2$ describes the quadratic spectrum of a particle in an infinite well. At sufficiently low temperatures, only the ground state and first excited state are thermally populated. Given that $\Delta = E_1 - E_0 = C$ and assuming  non-degenerate levels, the low-T QFI follows the universal form
\begin{equation}
	\mathcal{F}^{\text{low}}(T) \simeq \frac{C^2}{T^4} e^{-C/T}.
\end{equation}
At high temperatures, we distinguish two scenarios that correspond to finite (bounded) and infinite (unbounded) spectra. Firstly, when the temperature is much higher than the highest energy level, all $N$ states are equally populated. The QFI is then determined by the spectral variance
\begin{equation}
	\mathrm{Var}(H) = \frac{1}{N}\sum_{n=0}^{N-1} E_n^2 - \left(\frac{1}{N}\sum_{n=0}^{N-1} E_n\right)^2,
\end{equation}
as provided in Eq.~\eqref{eq:limit qfi infinite T}. For sake of simplicity, redefining the sums as
\begin{equation}
	S_1(\alpha,N) = \sum_{n=0}^{N-1} n^\alpha, \qquad
	S_2(\alpha,N) = \sum_{n=0}^{N-1} n^{2\alpha},
\end{equation}
for large $N$, we can perform the approximations
\begin{align}
	S_1 \simeq  \frac{N^{\alpha+1}}{\alpha+1}, \quad
	S_2 \simeq  \frac{N^{2\alpha+1}}{2\alpha+1},
\end{align}
leading to a QFI in the form 
\begin{equation}
    \label{eq:PL_high, finite}
	\mathcal{F}^{\text{high, finite}}(T) \simeq \frac{C^2 N^{2\alpha}}{T^4} \cdot \frac{\alpha^2}{(2\alpha+1)(\alpha+1)^2},
\end{equation}
valid for $N\gg1$ and $T\gg CN^\alpha$. Therefore, finite spectra exhibit the universal bounded-system scaling $\mathcal{F}\propto T^{-4}$, while the dependence on the spectral exponent $\alpha$ determines how thermometric sensitivity scales with the spectral span $E_{\max}\sim N^\alpha$. The scaling $N^{2\alpha}$ at fixed $C$ reflects the growth of the spectral bandwidth and is only applied within the high-temperature regime $T \gg E_{\max}$.

On the other hand, for an infinite number of states, the partition function at high $T$ is
\begin{align}
	Z \simeq \int_0^{\infty} e^{-C x^\alpha/T} dx
	= \frac{\Gamma(1/\alpha)}{\alpha} \left(\frac{T}{C}\right)^{1/\alpha},
\end{align}
where $\Gamma(x)$ denotes the Euler Gamma function with $ \Gamma(x)=\int_0^\infty t^{x-1}e^{-t}\,dt $. The mean energy is
\begin{align}
	\langle H \rangle = \frac{1}{Z} \frac{1}{\alpha \beta} (\beta C)^{-1/\alpha} \Gamma\left(1+\frac{1}{\alpha}\right) = \frac{T}{\alpha}.
\end{align}
Similarly, for $\langle H^2 \rangle$:
\begin{align}
	\langle H^2 \rangle \simeq \frac{T^2}{\alpha} \left(1 + \frac{1}{\alpha}\right).
\end{align}
Therefore, the QFI for the infinite spectrum at high temperature yields
\begin{equation}
    \label{eq:PL_high, infinite}
	\mathcal{F}^{\text{high, infinite}}(T) \sim \frac{1}{\alpha T^2}.
\end{equation}
For a finite number of levels and at fixed $C$, the prefactor of Eq.~\eqref{eq:PL_high, finite} grows as $N^{2\alpha}$. This growth is entirely due to the increase of the bandwidth $CN^\alpha$, and it is not an intrinsic advantage of rapidly growing spectra. For an unbounded spectrum, the QFI no longer depends on $N$ and decays more slowly with temperature.

\subsection{Quantum thermometry using systems with hybrid spectra}
\label{subsec:hybrid}

We now consider a class of quantum systems whose energy spectrum contains both discrete bound states and a continuous part. Such situations arise naturally in many physical systems~\cite{MiroshnichenkoRevModPhys2010,GarmonFotschPhys2013}. However, the canonical partition function can be written in the form
\begin{equation}
	Z(\beta) = \sum_{n} g_n e^{-\beta E_n} + \int_{E_c}^{\infty} dE\, \rho(E) e^{-\beta E},
\end{equation}
where $E_n$ and $g_n$ denote the discrete energy levels and their degeneracies, $\rho(E)$ is the density of states of the continuum, and $E_c$ is the threshold energy above which the spectrum becomes continuous. We now derive the asymptotic behavior of QFI in both regimes, low and high $T$. To proceed with the calculations, we assume that the system possesses a non-degenerate ground state $E_0$, and the first excitation gap is $\Delta = E_1 - E_0 > 0$. Additionally, without loss of generality, we consider that the continuum spectrum starts at $E_c > E_1$, with the additional condition that the discrete spectrum is completely below the continuum threshold $E_{c}$. The density of states near the threshold behaves as
\begin{equation}
    \label{eq:density of states hybrid}
	\rho(E) \simeq C (E - E_c)^s, \quad E \gtrsim E_c.
\end{equation}
\textcolor{black}{Throughout, the bound spectrum is finite, $N_d=\sum_n g_n<\infty$. We define the ground-state continuum gap $\Delta_c=E_c-E_0$.}
The QFI for this scenario separates into the following contributions
\begin{equation}
	\mathcal{F}(T)=
	\mathcal{F}_{d}(T)
	+\mathcal{F}_{c}(T)
	+\mathcal{F}_{dc}(T),
\end{equation}
where
\begin{equation}
	\mathcal{F}_{d}=\frac{p_d\,\mathrm{Var}_d}{T^4},
	\quad
	\mathcal{F}_{c}=\frac{p_c\,\mathrm{Var}_c}{T^4},
	\quad
	\mathcal{F}_{dc}=
	\frac{p_d p_c(\mu_d-\mu_c)^2}{T^4}.
\end{equation}
The QFI naturally separates into discrete, continuous, and mixed contributions (see appendix~\ref{discrete and continuous:AMD} for more details). The first two terms quantify thermal fluctuations within each spectral sector, whereas the mixed term arises from the difference between their mean energies and is relevant when both cases are appreciably populated.

In the low temperature limit $T \to 0$ ($\beta \to \infty$), the QFI takes the universal form
\begin{equation}\label{eq:hybrid_lowT}
	\mathcal F(T) \simeq \frac{g_1 \Delta^2}{T^4} \exp\!\left(-\frac{\Delta}{T}\right).
\end{equation}
\color{black}
This form depends only on $\Delta$ and $g_1$, and holds for $T\ll\Delta$ and as long as the continuum regime is negligible. If there is no bound excited level, and the first excitation is the continuum edge, one finds instead
\begin{equation}
	\label{eq:hybrid_lowT_cont}
	\mathcal F(T) = C\,\Gamma(s+1)\,\Delta_c^{2}\,T^{\,s-3}\,e^{-\Delta_c/T},
\end{equation}
this has the same $T^{-4}e^{-\text{gap}/T}$ structure relative to Eq.~\eqref{eq:hybrid_lowT}, but with the degeneracy $g_1$ replaced by the temperature-dependent effective degeneracy $C\Gamma(s+1)T^{s+1}$. In this case,  the threshold exponent $s$ already controls the low-$T$ law, whereas it is absent from Eq.~\eqref{eq:hybrid_lowT}.
\color{black}

In the opposite limit $T \to \infty$ ($\beta \to 0$), the discrete contribution becomes negligible compared to the continuum, and the QFI behaves as
\begin{equation}\label{eq:hybrid_highT}
	\mathcal F(T) \simeq \frac{s+1}{T^2}.
\end{equation}
with $s>-1$. The low-temperature limit is related to the gap $\Delta$, whereas the high-temperature limit requires the high-energy form of $\rho(E)$: we assume that the power law holds for all $E\ge E_c$ (as for a free particle) or at least asymptotically for $E\to\infty$ (see Appendix \ref{discrete and continuous:AMD}). The exponent $s$ characterizes the threshold behavior of the continuum density of states. For a free particle, $s=-1/2$ corresponds to a 1D system, $s=0$ to a 2D system (constant density of states), and $s=1/2$ to a 3D system.

\color{black}
Equation~\eqref{eq:hybrid_highT} simply implies that $C_V(T)\to s+1$. The bound states do not change or destroy this result but it add only small corrections to it. To keep things simple, we assume that the power law $\rho(E)=C(E-E_c)^s$ for all $E\ge E_c$. With this assumption, the continuum partition function yields $Z_c=C\,\Gamma(s+1)\,T^{s+1}e^{-E_c/T}$, and from it we get the continuum mean energy and variance as $\mu_c=E_c+(s+1)T$ and $\mathrm{Var}_c=(s+1)T^2$. For the high-temperature regime $T\gg E_c-E_0$, every bound level has $e^{-(E_n-E_c)/T}\simeq1$, therefore the ratio of the two partition functions yields
\begin{equation}
	\label{eq:hybrid_eps}
	\frac{p_d}{p_c}=\frac{Z_d}{Z_c}=\frac{\sum_n g_n e^{-(E_n-E_c)/T}}{C\,\Gamma(s+1)\,T^{s+1}}\simeq\epsilon(T)=\frac{N_d}{C\,\Gamma(s+1)\,T^{s+1}}.
\end{equation}
When $\epsilon\ll1$ it means that all statistical weight sits in the continuum and  $p_d\simeq\epsilon$ and $p_c\simeq1-\epsilon$. Therefore, inserting these weights in the decomposition, the continuum term is $\mathcal{F}_c\simeq(1-\epsilon)(s+1)/T^2$, the mixed term is $\mathcal{F}_{dc}\simeq\epsilon\,(s+1)^2/T^2$ because $\mu_c-\mu_d\simeq(s+1)T$, and $\mathcal{F}_d\le\epsilon\,(E_c-E_0)^2/T^4$ is negligible (see Eq.~\eqref{eq:Var_bound}). Summing the terms gives
\begin{equation}
	\label{eq:hybrid_highT_corr}
	\mathcal{F}(T)\simeq\frac{s+1}{T^{2}}(1+s\,\epsilon(T)).
\end{equation}
Both low- and high-temperature laws are controlled by different parameters, for instance at low $T$ by $\Delta$ and $g_1$, and at high $T$ by $s$. By contrast, the crossover comes from a competition between two low-temperature regimes, when $T\ll\Delta,\Delta_c$, one has $\mathrm{Var}(H)\simeq\sum_i p_i(E_i-E_0)^2$, so the level $E_1$ gives Eq.~\eqref{eq:hybrid_lowT} and the continuum gives Eq.~\eqref{eq:hybrid_lowT_cont}, with ratio
\begin{equation}\label{eq:hybrid_ratio}
	R(T)\simeq W\,T^{\,s+1}\,e^{-(E_c-E_1)/T},
	\quad W=\frac{C\,\Gamma(s+1)\,\Delta_c^{2}}{g_1\Delta^{2}}.
\end{equation}
Because $E_c>E_1$, the ratio goes to zero as $T\to0$, so at very low-temperature the bound Eq.~\eqref{eq:hybrid_lowT} is always the true low-temperature limit. Given the ratio $R$ increases temperature, therefore $R(T^{\star})=1$ defines a crossover temperature and we have
\begin{equation}\label{eq:hybrid_Tcross}
	T^\star=\frac{E_c-E_1}{\ln\big[W(T^\star)^{s+1}\big]},
\end{equation}
\color{black}

\subsection{Quantum thermometry using atomic and molecular probes}

Since the miniaturization of modern technologies has enabled the use of atomic or molecular systems such as quantum sensors \cite{xu2018molecular,demille2024quantum,du2024single}, we also focus on the thermometric performance of such physical models.

\subsubsection{Hydrogen-like systems} 
The most widely used model for describing atomic systems is the hydrogen-like atom, whose spectrum arises from the Coulomb interaction and can be systematically refined through fine-structure, hyperfine-structure, and Zeeman corrections. For the hydrogen-like atom, the full Hamiltonian reads
\begin{equation}
	H = H_{\mathrm{C}} + H_{\mathrm{fs}} + H_{\mathrm{hfs}},
\end{equation}
where each contribution is defined as
\begin{equation}
	H_{\mathrm{C}} = \frac{\mathbf{p}^2}{2m} - \frac{Z e^2}{4\pi\epsilon_0 r}=\frac{\mathbf{p}^2}{2m} + V(r).
\end{equation}
This Hamiltonian yields the well-known hydrogen spectrum
\begin{equation}
    \label{eq:hydrogen energies}
	E_n^{(0)} = -\frac{m Z^2 e^4}{2(4\pi\epsilon_0)^2}\frac{1}{n^2},
	\quad n=1,2,\dots
\end{equation}
Each energy level exhibits $n^2$ degeneracy (excluding the degree of freedom of the spin). Fine structure corrections, which arise from relativistic effects and spin-orbit coupling, are described by
\begin{equation}
	H_{\mathrm{fs}} = H_{\mathrm{rel}} + H_{\mathrm{SO}} + H_{\mathrm{D}},
\end{equation}
with
\begin{align}
	\begin{cases}
		H_{\mathrm{rel}} &= -\frac{\mathbf{p}^4}{8m^3c^2}, \\
		H_{\mathrm{SO}} &= \frac{1}{2m^2c^2}\frac{1}{r}\frac{dV}{dr}\,\mathbf{L}\cdot\mathbf{S}, \\
		H_{\mathrm{D}} &= \frac{1}{8m^2c^2}\nabla^2 V 
	\end{cases}.
\end{align}
The resulting spectrum of the Coulomb + fine structure Hamiltonian depends on the total angular momentum $j$ and reads
\begin{equation}
	E_{n j} = mc^2
	\left[
	1+\left(
	\frac{Z\alpha}{n-j-\frac12 + \sqrt{(j+\frac12)^2 - Z^2\alpha^2}}
	\right)^2
	\right]^{-1/2}.
\end{equation}
Where $ \alpha=\frac{e^2}{4\pi\epsilon_0 c} $. In the regime $Z\alpha \ll 1$, which corresponds to weak relativistic effects. The energy can be expanded perturbatively in powers of $(Z\alpha)^2$. Thus, the fine structure expression reads
\begin{equation}
	E_{n j} \simeq E_n^{(0)}
	\left[
	1+\frac{(Z\alpha)^2}{n}
	\left(
	\frac{1}{j+\frac12}-\frac{3}{4n}
	\right)
	\right].
\end{equation}
The hyperfine interaction between nuclear spin $\mathbf{I}$ and electronic angular momentum $\mathbf{J}$ is described by
\begin{equation}
	H_{\mathrm{hfs}} = A\,\mathbf{I}\cdot\mathbf{J}
	= \frac{A}{2}
	\left(
	\mathbf{F}^2 - \mathbf{I}^2 - \mathbf{J}^2
	\right),
\end{equation}
where $\mathbf{F}=\mathbf{I}+\mathbf{J}$ is the total angular momentum. The total Hamiltonian $H_{\mathrm{C}}+H_{\mathrm{fs}}+H_{\mathrm{hfs}}$ remains diagonal in the basis $\ket{n,j,F,m_F}$ and yields the spectrum
\begin{equation}
	E_{n j F} = E_{n j}
	+ \frac{A}{2}
	\left[
	F(F+1) - I(I+1) - J(J+1)
	\right],
\end{equation}
with
\begin{equation}
	F = |I-J|, \dots, I+J.
\end{equation}
Hydrogen-like atoms provide a natural platform for sensing in general since their energy spectra are well characterized and highly sensitive to external perturbations. Such systems offer a conceptually simple and broadly applicable framework for quantum thermometry. 
The energy spectrum for the total Hamiltonian reads
\begin{equation}
	E_{n j F} = E_n^{(0)} + \Delta_{\mathrm{fs}}(n,j) + \Delta_{\mathrm{hfs}}(n,j,F),
\end{equation}
with degeneracy;
\begin{equation}
	g_{n j F} = 2F+1.
\end{equation}
For this model, there is no specific simplification of the formula for the quantum Fisher information in terms of temperature. Therefore, for convenience, we refer to Eq.~\eqref{eq:(q)fi}. Nevertheless, in the low- and high-temperature limits, interesting trends emerge. At sufficiently low temperatures, only the ground-state hyperfine manifold is populated. For hydrogen, with $n=1$ and $j=1/2$, the hyperfine splitting between $F=0$ and $F=1$ is $\Delta_{\mathrm{hfs}} = A$, with degeneracies $g_0 = 1$ and $g_1 = 3$. The QFI for this effective two-level system reads as
\begin{equation}
	\mathcal{F}_T^{\mathrm{low}}(T) \simeq 
	\frac{3A^2}{T^4} e^{-A/T}.
\end{equation}
At high temperature, the Boltzmann factors can be expanded as
\begin{equation}
	e^{-\beta E_n} \simeq 1 - \beta E_n + \mathcal{O}(\beta^2).
\end{equation}
In this regime, fine and hyperfine splittings become negligible, and the spectrum can be approximated by the hydrogenic energies of Eq.~\eqref{eq:hydrogen energies} as
\begin{equation}
	E_n = -\frac{E_0}{n^2}, \quad g_n = \gamma n^2, \quad n=1,\dots,N,
\end{equation}
where $\gamma$ accounts for internal degeneracies. 
In an isolated atom, the bound levels accumulate at the ionization threshold ($n\to\infty$), so a finite $N$ is not a property of the atom: it must be fixed by external conditions, or by an approximated regime. The expressions below are therefore results for a ``truncated hydrogenic model''.
As a result, the partition function becomes
\begin{equation}
	Z \simeq \sum_{n=1}^N g_n = \gamma \frac{N(N+1)(2N+1)}{6}.
\end{equation}
For large $N$, one obtains
\begin{align}
	\langle H \rangle \simeq -\frac{3E_0}{N^2}, \quad
	\langle H^2 \rangle \simeq \frac{E_0^2 \pi^2}{2N^3}.
\end{align}
Thus, the variance yields the following
\begin{equation}
	\mathrm{Var}(H) \simeq \frac{E_0^2 \pi^2}{2N^3}.
\end{equation}
Finally, the QFI reads
\begin{equation}
	\mathcal{F}_T^{\mathrm{high}}(N,T)
	\simeq
	\frac{E_0^2 \pi^2}{2}
	\frac{1}{T^4 N^3}.
\end{equation}
This behavior is analogous to that of finite-level models with degeneracies Eq.~\eqref{eq:high_T_QFI_atom_degenerate}, where the high-temperature sensitivity is determined not by the number of available states, but by the rate at which energy levels become thermally indistinguishable. In hydrogenic spectra, this effect is governed by the hierarchy of progressively decreasing level spacings $ \Delta E_n = E_0 \frac{2n+1}{n^2(n+1)^2} $. As $N$ increases, higher Rydberg states accumulate and $ \Delta E_n \sim n^{-3} \to 0 $ so that these levels contribute predominantly to normalization rather than to energy fluctuations. In this sense, increasing the spectral manifold does not enhance precision, on the contrary, it weakens the sensitivity.

\subsubsection{Diatomic Molecules} 
Combining the effects of electrostatic potential, rotational, and vibrational modes, it is possible to describe diatomic molecular systems, including homonuclear molecules composed of light elements and heteronuclear molecules such as carbon monoxide and nitric oxide. The Hamiltonian of such systems may be written as
\begin{equation}
    \label{eq:M}
    H^{m} = H^{eq} + H^{rm} + H^{vm},
\end{equation}
where the three terms represent the equilibrium Hamiltonian and the rotational and vibrational mode Hamiltonians, respectively. The vibrational potential may be conveniently represented using the Morse potential as
\begin{equation}
    \label{eq:Mvm}
    V^{vm} = D_e (1-e^{-a(r-r_{eq})})^2,
\end{equation}
where $r$ and $r_{eq}$ represent the distance between the atoms of the molecule and the equilibrium bond distance, respectively; $D_e$ is the energy necessary to break the bond of the molecule; and $a$ controls the width of the potential. The energy levels of such a Hamiltonian are
\begin{equation}
    \label{eq:eMvm}
    E^{vm}_{n} \simeq \omega \left( n +\frac{1}{2} \right) - \omega \chi \left( n +\frac{1}{2} \right)^2,
\end{equation}
with $\chi= a^2 / (2\mu \omega)$ and $\mu$ is the reduced mass of the molecule, corresponding to the sum of the energy levels of a harmonic oscillator together with an anharmonic correction. The Morse potential supports only a finite number of bound vibrational levels, with decreasing spacing, followed by a continuum. Therefore, a diatomic molecule has a hybrid spectrum (see Sec.~\ref{subsec:hybrid}). The discussion below is qualitative and refers to the discrete part of the spectrum, where rotational and vibrational levels are populated sequentially. In the same way, the rotational energy levels correspond to the correction of the quantum rotor energy levels, considering asymmetrical factors as
\begin{align}
    \label{eq:eMrm}
    E^{\rm rm}_n\simeq B\,n(n+1)-D\,n^2(n+1)^2,
\end{align}
with $D=\frac{4B^3}{\omega^2}=\frac{1}{2\mu^2kR_0^6}$ and $\omega^2=\frac{k}{\mu}$. Here, $\mu$ is the reduced mass, $k$ the harmonic force constant, and $R_0$ the internuclear distance. The factor $B = 1/(2\mu R_0^2)$, also called rotational constant, sets the energy scale for rotational transitions~\cite{brown2003rotational}.

The thermometric properties of diatomic molecules can be naturally understood using the fact that their spectrum decomposes into rotational and vibrational sectors, the rotational levels $E_n^{\mathrm{rot}} = B n(n+1)$ correspond to the quantum rotor model, while the vibrational spectrum  $E_n^{\mathrm{vib}} \simeq \omega (n+1/2) - \chi\omega (n+1/2)^2$ describes a weakly anharmonic oscillator. 

As for Hydrogen-like systems, also for this model, there is no specific simplification of the formula for the quantum Fisher information of the temperature beyond Eq.~\eqref{eq:(q)fi}.
However, this structure leads to a clear hierarchy of energy scales, with $B \ll \omega$ in typical molecules~\cite{CalderJCPDiatomic,KuncJPCDiatomic}. Therefore, the low-temperature response is governed by the smallest excitation gap associated with rotational transitions. In this regime, the molecule effectively behaves as a few-level system and exhibits universal scaling $\mathcal{F}(T) \sim (g_1 \Delta^2/T^4)e^{-\Delta/T}$, with $\Delta \sim 2 B$ and rotational degeneracy $g_1=3$.

As temperature increases, rotational states become thermally populated before vibrational excitations contribute appreciably~\footnote{Because $B \ll \omega$, rotational excitations enter the thermal window at significantly lower temperatures than vibrational ones.} producing a gradual crossover from discrete-level thermodynamics to a regime where the rotational spectrum becomes thermally dense and thermodynamic quantities admit an effective continuum approximation. In the high temperature regime, the rotational contribution recovers the characteristic $\mathcal{F}(T)\propto 1/T^2$ scaling of a rotor-like spectrum, the vibrational mode adds its own contribution, while the anharmonicity contributes only sub-leading corrections. Diatomic molecules, therefore, provide a physical realization of the crossover between the discrete and continuous thermometric regimes.

\subsection{Quantum thermometry with free particles}
\label{subsec:Qtwfp}

As a final model, after analyzing paradigmatic discrete finite and infinite discrete (or mixed) spectra, we consider the free-particle model in the continuous-spectrum limit in spatial dimensions $d=1,~2,~3$. The Hamiltonian of such a system is defined as
\begin{equation}
    \label{eq:H_free}
    H^{free} = \sum_{k} \frac{p_{k}^{2}}{2m} = E^{free}.
\end{equation}

The energy probability density
\begin{equation}
	p_T(E)=\frac{g(E)\,e^{-\beta E}}{Z(T)},\
\end{equation}
and the partition function reads
\begin{equation}
	Z(T)=\int_0^\infty g(E)\,e^{-\beta E}\,dE,
\end{equation}
where $g(E)$ is the density of states. The QFI is  determined by the energy fluctuations,
\begin{align}
	\mathcal{F}_T
	&= \int_0^\infty p_T(E)\left[\frac{\partial \ln p_T(E)}{\partial T}\right]^2 dE,
\end{align}
where $\langle E\rangle=\int_0^\infty E\,p_T(E)\,dE$.
For a single free particle in a three-dimensional volume $V$, the single-particle density of states is $g_1(E)=\frac{V}{4\pi^2}\!\left(2m\right)^{3/2}\!\sqrt{E}$. Using the canonical averages for the free-particle spectrum yields
\begin{equation}
	\langle E\rangle_1=\frac{3}{2}T,
	\qquad
	\mathrm{Var}_1(E)=\frac{3}{2}T^2.
\end{equation}
Hence, the single-particle QFI is
\begin{equation}
	\mathcal{F}_1(T)=\frac{\mathrm{Var}_1(E)}{T^4}=\frac{3}{2T^2}.
\end{equation}
For $N$ non-interacting particles 
\begin{equation}
	\mathcal{F}_N(T)=\frac{N\,\mathrm{Var}_1(E)}{T^4}=\frac{3N}{2T^2}.
\end{equation}    
This result demonstrates the $T^{-2}$ scaling of scale-free unbounded spectra, with prefactor $d/2$ per particle and linear scaling with $N$, thereby establishing the free particle as a benchmark for continuous-spectrum thermometry.

\begin{table*}[t]
	\centering
	\caption{Spectral properties and asymptotic QFI scaling for the thermometric models considered in this work. \textcolor{black}{The entries for the $N$-level atom/complete graph and for the ring are also discussed in Refs.\cite{correa2015individual, candeloro2021role}; all other entries are obtained by specializing Eqs.\eqref{eq:(q)fi}-\eqref{eq:limit qfi infinite T}.}}
	\renewcommand{\arraystretch}{1.6}
	\begin{tabular}{|l|c|c|c|c|c|c|}
		\hline
		\textbf{Model} & \textbf{Spectrum } $ E_n $ & \textbf{Degeneracy } $ g_n $ & \textbf{Gap}  & \textbf{Dimension} & \textbf{QFI low  T} & \textbf{QFI high T} \\
		\hline
		
		Spin system 
		& $-\omega m$
		& $1$
		& $\omega$
		& $ N = 2s+1 $
		& $\frac{\omega^2}{T^4} e^{-\omega/T}$
		& $\frac{\omega^2(N^2-1)}{12 T^4}$ \\
		
		\hline
		
		$N$-level atom/QW
		& $0,\epsilon$
		& $1, N-1$
		& $\epsilon$
		& N 
		& $\frac{(N-1)\epsilon^2}{T^4} e^{-\epsilon/T}$
		& $\frac{(N-1)\epsilon^2}{N^2 T^4}$\\
		
		\hline
		
		Ring topology (QW)
		& $2\omega-2\mathcal{J}\cos\left(\frac{2\pi(l+1)}{N}\right)$
		& $g_0=1$, $g_{n\ge1}=2$
		& $\frac{4\pi^2\mathcal{J}}{N^2}$
		& N 
		& $\frac{32\pi^4\mathcal{J}^2}{N^4T^4} e^{-4\pi^2\mathcal{J}/(N^2T)} $
		& $\frac{2\mathcal{J}^2}{T^4}$\\
		
		\hline
		
		Quantum well (1D)
		& $\alpha n^2$
		& $1$
		& $ 3 \alpha $
		& Infinite 
		& $\frac{9\alpha^2}{T^4} e^{-3\alpha/T}$
		& $\frac{1}{2T^2}$\\
		
		\hline
		
		Harmonic oscillator
		& $\omega(n+1/2)$
		& $1$
		& $\omega$
		& Infinite 
		& $\frac{\omega^2}{T^4} e^{-\omega/T}$
		& $\frac{1}{T^2}$\\
		
		\hline
		
		Quantum rotor
		& $B n(n+1)$
		& $2n+1$
		& $ 2 B $
		& Infinite 
		& $\frac{3 (2B)^2}{T^4} e^{-2B/T}$
		& $\frac{1}{T^2}$\\
		
		\hline
		
		Power-law spectrum
		& $C n^\alpha$
		& $1$
		& $ C $
		& Finite $ N $ 
		& $\frac{C^2}{T^4} e^{-C/T}$
		& $\frac{C^2 N^{2\alpha}}{T^4} \frac{\alpha^2}{(2\alpha+1)(\alpha+1)^2}$\\
		
		\hline
		
		Power-law spectrum
		& $C n^\alpha$
		& $1$
		& $ C $
		& $N \rightarrow\infty$ 
		& $\frac{C^2}{T^4} e^{-C/T}$
		& $\frac{1}{\alpha T^2}$\\
		
		\hline
		
		Hydrogen-like atom
		& $-E_0/n^2$
		& $\gamma n^2$
		& $ A $
		& Finite 
		& $\frac{3 A^2}{T^4} e^{-A/T}$
		& $\frac{E_0^2\pi^2}{2T^4 N^3}$\\
		
		\hline
		
		Free particle (3D)
		& $p^2/2m$
		& ---
		& ---
		& Continuous 
		& $\frac{3}{2T^2}$
		& $\frac{3}{2T^2}$\\
		
		\hline
		
		Hybrid spectrum
		& $\{E_n\}\cup [E_c,\infty)$
		& $ \rho(E)\sim(E-E_c)^s$
		& $\Delta$
		& Mixed
		& $\frac{g_1\Delta^2}{T^4}e^{-\Delta/T}$
		& $\frac{s+1}{T^2}$ \\
		\hline
	\end{tabular}
	\label{tab:unified_models}
\end{table*}

\section{Summary and discussion}
\label{sec:S}
In this Section, we summarize and discuss the results obtained throughout Section \ref{sec:R}. Specifically, Table~\ref{tab:unified_models} collects the characteristics of each physical system considered: energy spectrum, degeneracy, relevant energy gaps, and QFI limits. The asymptotic scalings reveal universal structures that transcend specific physical realizations, as reported in Table~\ref{tab:unified_models}. At low temperatures, every gapped probe with a non-degenerate ground state and $g_1e^{-\Delta/T}\ll1$ exhibits the same functional dependence:
\begin{equation}
	\mathcal{F}(T) \sim \frac{g_1 \Delta^2}{T^4} \, e^{-\Delta / T},
\end{equation}
with $g_1$, which is the degeneracy of the energy level $E_1$ and $\Delta=E_1 - E_0$. This universal form arises because only the ground state and the first excited level are thermally populated when $T \ll \Delta$. Consequently, optimizing low-temperature sensitivity for a fixed gap reduces to \emph{maximizing the degeneracy $g_1$} of the first excited state (see Sec.~\ref{sec:R} or Ref.~\cite{correa2015individual} for details). Although the quantum rotor and the hydrogen hyperfine transition each offer $g_1 = 3$, providing a factor-of-three enhancement over a non-degenerate two-level system, substantially larger degeneracies are achievable. Quantum walks on a fully connected graph realize an effective Hamiltonian whose spectrum is exactly that of an $N$-level atom, i.e., a non-degenerate ground state and an $(N-1)$-fold degenerate excited manifold. For such a system, the low-temperature QFI becomes
\begin{equation}
	\mathcal{F}^{N}(T) \simeq \frac{(N-1)\Delta^2}{T^4} \, e^{-\Delta / T},
\end{equation}
which scales linearly with the number of nodes $N$. Thus, by engineering quantum walks on sufficiently large fully connected networks, one can, in principle, achieve a large low-temperature sensitivity enhancement, until the condition $(N-1)e^{-\Delta/T}\simeq 1$ is reached, beyond which further increases in $N$ become destructive. For any discrete spectrum, no physical system can evade fundamental exponential suppression with $T$; it can only be compensated by a degeneracy growing as $N-1\gtrsim e^{\Delta/T}$, i.e., exponentially large in $\Delta/T$. Although QW on fully connected graphs exploits degeneracy to enhance low-temperature sensitivity, an alternative mechanism emerges in the ring topology, where the excitation gap becomes the primary tunable resource. The low-temperature QFI of a QW in a ring topology reveals competition; increasing $N$ reduces the first energy gap $\Delta \simeq 4\pi^2\mathcal{J}/N^2$, enhancing thermal accessibility, but suppresses the prefactor as $N^{-4}$. 
Unlike fully connected graphs, where the enhancement arises from degeneracy, the ring takes advantage of spectral compression: increasing $N$ continuously reduces the gap, allowing the probe to adapt to its optimal operating temperature. This identifies gap engineering as an alternative resource for low-temperature quantum thermometry.

At high temperatures, the behavior can be categorized into two distinct universality classes, depending on whether the energy spectrum is bounded or unbounded and scale-free. For systems with a finite, bounded spectrum, the QFI decays as $\mathcal{F}(T) \sim T^{-4}$. However, the prefactor exhibits a strong dependence on the size of the system and the spectral exponent $\alpha$, where $E_n = C n^{\alpha}$, so that the QFI scales as
\begin{equation}
	\mathcal{F}^{\text{high, finite}}(T) \sim \frac{C^2 N^{2\alpha}}{T^4} \cdot \frac{\alpha^2}{(2\alpha+1)(\alpha+1)^2}.
\end{equation}
At fixed $C$, the prefactor grows polynomially with $N$, but only because the bandwidth $CN^\alpha$ grows; at fixed bandwidth, rapidly growing spectra offer no intrinsic advantage.

In contrast, systems with an infinite unbounded spectrum exhibit a markedly different high-temperature scaling: $\mathcal{F}(T) \sim T^{-2}$, which is a significantly slower decay with temperature. For a continuous set of modes with a power-law density of states $g(\omega) = A \omega^s$, the high-temperature QFI scales as $\mathcal{F}(T) \sim (s+1)/T^2$. The free particle in three dimensions corresponds to $s=1/2$, and achieves the largest prefactor among this class, with $\mathcal{F}(T) = 3/(2T^2)$. 

Comparisons between probes are meaningful only once the constrained resources are specified. At fixed bandwidth, finite-bounded spectrum systems offer prefactors that scale as $N^{2\alpha}$, enabling large enhancements at the cost of a $T^{-4}$ decay, whereas infinite-spectrum systems provide a slower $T^{-2}$ decay but lack the possibility of size scaling. No single model is universally optimal. The appropriate thermometer is the one suited to the target temperature range and physical constraints.


\section{Conclusions}
\label{sec:C}

In conclusion, we have addressed the problem of temperature estimation by quantum probes from a spectral perspective, directly linking the performance of a quantum thermometer to the distribution of its energy levels. Starting from the identity $\mathcal{F}=(\Delta H)^{2}/T^4=C_V/T^2$, we obtained exact expressions and asymptotic laws for a broad class of probes, from finite non-degenerate systems to degenerate, continuous, hybrid, atomic and molecular spectra, stating the regime of validity of each result. 
At high temperature, bounded spectra exhibit a $T^{-4}$ decay, whereas scale-free unbounded spectra achieve a slower $T^{-2}$ scaling; continuity or unboundedness alone is not sufficient. At low temperatures, the exponential suppression of sensitivity is universal for systems with fixed spectral gaps, but we have demonstrated that the prefactor can be enhanced by engineering degenerate excited states or by implementing a quantum walk on a fully connected network topology as a thermometer. Beyond these scaling results, we have also identified a concrete physical implementation for the optimal spectral configuration proposed in \cite{correa2015individual}, which remains optimal at any temperature provided that the coupling is re-tuned accordingly. Furthermore, quantum walks on a ring topology offer a complementary strategy: by tuning the system size, the exponential suppression typical of discrete-spectrum thermometry can be converted into a power-law scaling of $T^{-2}$. This gap-engineering approach offers a distinct low-temperature enhancement mechanism that does not rely on degeneracy, thereby constituting an additional resource for low-temperature quantum thermometry. In general, our findings deepen the connection between spectral properties and thermometric precision, showing that the degeneracy, spectral growth, and boundedness of the spectrum are not just details but active resources. Using these features, one can overcome limitations of equilibrium thermometry and design application-specific quantum thermometers with tailored sensitivity.

\begin{acknowledgments}
This work has been done under the auspices of GNFM-INdAM.
\end{acknowledgments}

\appendix

\section{Additional details}
\label{a:AMD}

In this Appendix, we report the details of the derivation of the energy variance (QFI) and its high- and low-temperature limits for specific models. The complete derivation for relatively straightforward models is presented in the main text.

\subsection{Spin Systems}
\label{SS:AMD}
The low- and high-temperature limits of the QFI for spin systems are based on the asymptotic expansions of the hyperbolic cosecant. In particular,
\begin{align}
	\mathrm{csch}(x) &\simeq 2 e^{-x}, \quad x \gg 1, \nonumber \\
	\mathrm{csch}(x) &\simeq \frac{1}{x} - \frac{x}{6}, \quad x \ll 1.
\end{align}
For $T \to 0$ one has
\begin{equation}
	x = \frac{\omega}{2 T} \gg 1,
\end{equation}
so that
\begin{equation}
	\mathrm{csch}^2(x) \simeq 4 e^{-2x}.
\end{equation}
Therefore,
\begin{align}
	\mathrm{csch}^2\!\left(\frac{\omega}{2 T}\right)
	&\simeq
	4 e^{-\omega/T}, \nonumber \\
	\mathrm{csch}^2\!\left(\frac{N\omega}{2 T}\right)
	&\simeq
	4 e^{-N\omega/T}.
\end{align}
Since $e^{-N\omega/T} \ll e^{-\omega/T}$ for $N>1$, the second term is negligible and we obtain
\begin{equation}
	\mathrm{Var}(H^{s})
	\simeq
	{\omega^2} e^{-\omega/T}.
\end{equation}
Thus, at low temperature, the energy variance depends only on the lowest energy gap.
For $T \to \infty$ we have
\begin{equation}
    x = \frac{\omega}{2 T} \ll 1,
\end{equation}
and
\begin{equation}
	\mathrm{csch}^2(x)
	\simeq
	\frac{1}{x^2} - \frac{1}{3}.
\end{equation}
Hence,
\begin{align}
	\mathrm{csch}^2\!\left(\frac{\omega}{2 T}\right)
	&\simeq
	\frac{4 T^2}{\omega^2}
	- \frac{1}{3}, \nonumber \\
	\mathrm{csch}^2\!\left(\frac{N\omega}{2T}\right)
	&\simeq
	\frac{4 T^2}{N^2 \omega^2}
	- \frac{1}{3}.
\end{align}
Therefore, when $ T \to \infty $ the variance is
\begin{equation}
	\mathrm{Var}(H^{s})
	\simeq
	\frac{\omega^2}{12}
	\left(N^2 - 1\right).
\end{equation}

\subsection{Quantum Rotor}
\label{QR:AMD}

Given that the quantum rotor model exhibits degeneracy $ g_n $, the canonical partition function yields
\begin{equation}
	Z(\beta) = \sum_{n=0}^{\infty} g_n e^{-\beta E^{qr}_n} \, .
\end{equation}
The energy variance of a quantum rotor is
\begin{align}
	\label{eq:variance rotor}
	Var(H^{qr}) = \left( \frac{1}{2I} \right)^2 &\bigg[ \frac{\sum_{n} g_n h_n^2 e^{-\beta E^{qr}_{n} } }{\sum_{n} g_n e^{-\beta E^{qr}_{n}} } - \nonumber \\
	&\left( \frac{\sum_{n} g_n h_n e^{-\beta E^{qr}_{n} } }{\sum_{n} g_n e^{-\beta E^{qr}_{n}} } \right)^2 \bigg],
\end{align}
with $h_n = n(n + 1)$. Accordingly, the QFI is
\begin{equation}
	\label{eq:qfi quantum rotor}
	\mathcal{F}^{qr}(T) = \frac{\mathrm{Var}(H^{qr})}{T^4}.
\end{equation}
However, in the low temperature limit $T \to 0$ ($\beta \to \infty$), thermal occupation is effectively restricted to the lowest energy levels. The ground state $n=0$ is characterized by energy $E_0 = 0$ and degeneracy $g_0 = 1$. The first excited state ($n=1$) has energy
\begin{equation}
	E_1 = 2 \frac{1}{2I} = 2B,
\end{equation}
degeneracy $g_1 = 3$, and energy gap
\begin{equation}
	\Delta = E_1 - E_0 = 2B = \frac{1}{I}.
\end{equation}
At sufficiently low temperatures, one can approximate the system as a two-level system consisting of a non-degenerate ground state and a triply degenerate excited state. The partition function in this case is
\begin{equation}
	Z \simeq 1 + 3 e^{-\beta\Delta}.
\end{equation}
The mean energy is
\begin{equation}
	\langle H \rangle \simeq \frac{3\Delta e^{-\beta\Delta}}{1 + 3 e^{-\beta\Delta}}.
\end{equation}
The mean square energy is
\begin{equation}
	\langle H^2 \rangle \simeq \frac{3\Delta^2 e^{-\beta\Delta}}{1 + 3 e^{-\beta\Delta}}.
\end{equation}
Therefore, the variance becomes
\begin{equation}
	\mathrm{Var}(H) = \langle H^2 \rangle - \langle H \rangle^2
	= \frac{3\Delta^2 e^{-\beta\Delta}}{(1 + 3 e^{-\beta\Delta})^2}.
\end{equation}
Given that $T \to 0$, then $e^{-\beta\Delta} \ll 1$, hence
\begin{equation}
	\mathrm{Var}(H) \simeq 3\Delta^2 e^{-\beta\Delta} = 3\Delta^2 e^{-\Delta/T}.
\end{equation}
The low-temperature approximation is
\begin{equation}
	\mathcal{F}(T) = \frac{\mathrm{Var}(H)}{T^4} \simeq \frac{3\Delta^2}{T^4} e^{-\Delta/T},
\end{equation}
with the degeneracy factor $g=3$ contributing an extra factor of $3$ in the prefactor.\\

In the high-temperature regime, namely $\beta \to 0$, a large number of energy levels contribute, and the summation can be approximated by an integral. Moreover, for large $n$ one has
\begin{equation}
	n(n+1) \simeq n^2,
	\qquad
	2n+1 \simeq 2n,
\end{equation}
therefore, the canonical partition function is
\begin{equation}
	Z \simeq \int_0^\infty 2n e^{-\beta B n^2} dn = \frac{1}{\beta B}.
\end{equation}
The mean energy is
\begin{equation}
	\langle H \rangle = - \partial_\beta \ln Z = \frac{1}{\beta} = T,
\end{equation}
and the variance becomes
\begin{equation}
	\mathrm{Var}(H) = \partial_\beta^2 \ln Z = \frac{1}{\beta^2} = T^2.
\end{equation}
Therefore the QFI at high T scales as
\begin{equation}
	\mathcal F(T) \sim \frac{T^2}{T^4} = \frac{1}{T^2}.
\end{equation}

\subsection{Quantum Wells}
\label{QW:AMD}

The spectrum of quantum wells has a level spacing that grows linearly with $ n $, i.e., $ \Delta E_n = E_{n+1} - E_n = \alpha (2n + 1) $. In this case, the partition function is
\begin{equation}
	Z(\beta) = \sum_{n=1}^\infty e^{-\beta E^{qw}_n} = \sum_{n=1}^\infty e^{-\alpha \beta n^2} \, .
\end{equation}
The above expression does not allow for a compact formula without summation; thus, one can introduce the Jacobi theta function as
\begin{equation}
	\vartheta_3(0, q) = \sum_{n=-\infty}^\infty q^{n^2}, \quad q = e^{-\alpha\beta},
\end{equation}
and then
\begin{equation}
	\vartheta_3(0, e^{-\alpha\beta}) = \sum_{n=-\infty}^\infty e^{-\alpha\beta n^2} = 1 + 2\sum_{n=1}^\infty e^{-\alpha\beta n^2}.
\end{equation}
Therefore, the partition function becomes
\begin{equation}
	Z(\beta) = \frac{1}{2} \left[ \vartheta_3(0, e^{-\alpha\beta}) - 1 \right].
\end{equation}
The variance can be written as a function of the Jacobi theta function as
\begin{equation}
	\mathrm{Var}(H^{qw}) = \frac{ \vartheta_3''(0,e^{-\alpha \beta}) }{\vartheta_3(0,e^{-\alpha \beta}) - 1}
	- \left( \frac{ \vartheta_3'(0,e^{-\alpha \beta}) }{\vartheta_3(0,e^{-\alpha \beta}) - 1} \right)^2,
\end{equation}
with $\vartheta_3'(0,e^{-\alpha \beta})$ and $\vartheta_3''(0,e^{-\alpha \beta})$ are the first and second derivatives of the Jacobi theta function with respect to $\beta$, respectively. Thus, the QFI is expressed as
\begin{equation}
	\mathcal{F}^{qw}(T) = \frac{\mathrm{Var}(H^{qw})}{T^4}.
\end{equation}
The resulting QFI formula, expressed through Jacobi functions $\vartheta_3$, is exact and universally valid for any system with a quadratic spectrum $E_n \propto n^2$.

For the low temperature $T \to 0$ approximation, we have $\beta \to \infty$, hence $q = e^{-\alpha\beta} \to 0$. Following that, the theta function expands as
\begin{equation}
	\vartheta_3(0,q) = 1 + 2q + 2q^4 + 2q^9 + \dots
\end{equation}
Thus,
\begin{equation}
	\vartheta_3(0,q) - 1 = 2q + 2q^4 + \dots
\end{equation}
The derivatives with respect to $\beta$ of the theta function read
\begin{eqnarray}
	\vartheta_3'(0,q) &=& -\alpha \sum_{n=-\infty}^\infty n^2 q^{n^2}, \nonumber \\
	\vartheta_3''(0,q) &=& \alpha^2 \sum_{n=-\infty}^\infty n^4 q^{n^2},
\end{eqnarray}
expand to leading order in $q$
\begin{eqnarray}
	\vartheta_3'(0,q) &\simeq& -\alpha\left( 2q + 8q^4 + \dots \right), \nonumber \\
	\vartheta_3''(0,q) &\simeq& \alpha^2\left( 2q + 32q^4 + \dots \right).
\end{eqnarray}
The first term in $\mathrm{Var}(H)$ can be approximated as
\begin{equation}
	\frac{\vartheta_3''}{\vartheta_3 - 1} \simeq \alpha^2 \frac{1 + 16q^3}{1 + q^3} \simeq \alpha^2(1 + 15q^3).
\end{equation}
The second term becomes
\begin{equation}
	\left( \frac{\vartheta_3'}{\vartheta_3 - 1} \right)^2
	\simeq \alpha^2 \left( \frac{1 + 4q^3}{1 + q^3} \right)^2 \simeq \alpha^2 (1 + 6q^3).
\end{equation}
Thus, we have
\begin{equation}
	\mathrm{Var}(H^{qw}) \simeq \alpha^2(1 + 15q^3 - 1 - 6q^3) = 9\alpha^2 q^3.
\end{equation}
Given that $q = e^{-\alpha/T}$, then
\begin{equation}
	\mathrm{Var}(H^{qw}) \simeq 9\alpha^2 e^{-3\alpha/T}.
\end{equation}
Finally, the QFI for $ T \to 0 $ reads
\begin{equation}
	\mathcal{F}^{qw}(T) \simeq \frac{9\alpha^2}{T^4} e^{-3\alpha/T}.
\end{equation}
The energy gap between the ground and first excited state is $E_2 - E_1 = 3\alpha$. The low-temperature scaling follows the universal form
\begin{equation}
	\mathcal{F}^{qw}(T) \sim \frac{\Delta^2}{T^4} e^{-\Delta/T},
\end{equation}
where $\Delta = 3\alpha$.
For $T \to \infty$, i.e., $\beta \to 0$, we have $q \to 1$. Let assume that $\delta = \alpha\beta \to 0$. Using the Jacobi  transformation~\cite{whittaker2020course}, we have
\begin{equation}
	\vartheta_3(0, e^{-\delta}) = \sqrt{\frac{\pi}{\delta}} \, \vartheta_3(0, e^{-\pi^2/\delta}).
\end{equation}
For small $\delta$, we have $\vartheta_3(0, e^{-\pi^2/\delta}) \to 1$, then
\begin{equation}
	\vartheta_3(0,q) \simeq \sqrt{\frac{\pi}{\delta}}.
\end{equation}
Writing $f(\delta) = \vartheta_3(0, e^{-\delta}) \simeq \sqrt{\pi} \delta^{-1/2}$, the derivatives are
\begin{eqnarray}
	f'(\delta) &\simeq& -\frac12 \sqrt{\pi} \delta^{-3/2}, \nonumber \\
	f''(\delta) &\simeq& \frac34 \sqrt{\pi} \delta^{-5/2}.
\end{eqnarray}	
Recall that $\vartheta_3'(0,q) = -\alpha f'(\delta)$, $\vartheta_3''(0,q) = \alpha^2 f''(\delta)$. Therefore, the variance reads
\begin{equation}
	\mathrm{Var}(H^{qw}) \simeq \frac{\alpha^2}{2} \delta^{-2}.
\end{equation}
Given that $\delta = \alpha/T$, then $\delta^{-2} = \frac{T^2}{\alpha^2}$, hence
\begin{equation}
	\mathrm{Var}(H^{qw}) \simeq \frac{T^2}{2}.
\end{equation}
Finally, the QFI at high temperature approximation reads
\begin{equation}
	\mathcal{F}^{qw}(T) \simeq \frac{1}{2T^2}.
\end{equation}

\subsection{Systems with both discrete and continuous spectra}
\label{discrete and continuous:AMD}

It is convenient to separate discrete and continuous contributions by defining
\begin{equation}
	Z(\beta) = Z_d(\beta) + Z_c(\beta),
\end{equation}
where
\begin{align}
	Z_d(\beta) &= \sum_{n} g_n e^{-\beta E_n},\\
	Z_c(\beta) &= \int_{E_c}^{\infty} dE \, \rho(E) e^{-\beta E}.
\end{align}
The mean energy is
\begin{equation}
	\langle H \rangle = \frac{\mathcal{M}_1^{(d)} + \mathcal{M}_1^{(c)}}{Z_d + Z_c},
\end{equation}
where
\begin{equation}
	\mathcal{M}_k^{(d)} = \sum_n g_n E_n^k e^{-\beta E_n}, \quad
	\mathcal{M}_k^{(c)} = \int_{E_c}^{\infty} dE \, \rho(E) E^k e^{-\beta E}.
\end{equation}
Similarly, we have
\begin{equation}
	\langle H^2 \rangle = \frac{\mathcal{M}_2^{(d)} + \mathcal{M}_2^{(c)}}{Z_d + Z_c}.
\end{equation}
The energy variance is therefore
\begin{equation}\label{eq:Var_d_c}
	\mathrm{Var}(H) 
	= \frac{\mathcal{M}_2^{(d)} + \mathcal{M}_2^{(c)}}{Z_d + Z_c}
	- \left( \frac{\mathcal{M}_1^{(d)} + \mathcal{M}_1^{(c)}}{Z_d + Z_c} \right)^2.
\end{equation}
The variance admits a more transparent form upon introducing
\begin{equation}
	p_d = \frac{Z_d}{Z_d + Z_c}, \qquad p_c = \frac{Z_c}{Z_d + Z_c},
\end{equation}
which satisfy $p_d + p_c = 1$ and give the equilibrium weights of the discrete and continuum sectors, respectively. Introducing the mean energies within the discrete and continuous
\begin{equation}
	\mu_d = \frac{\mathcal{M}_1^{(d)}}{Z_d}, \qquad \mu_c = \frac{\mathcal{M}_1^{(c)}}{Z_c},
\end{equation}
and the corresponding variances
\begin{equation}
	\mathrm{Var}_d = \frac{\mathcal{M}_2^{(d)}}{Z_d} - \mu_d^2, \qquad
	\mathrm{Var}_c = \frac{\mathcal{M}_2^{(c)}}{Z_c} - \mu_c^2.
\end{equation}
Substituting these definitions into the expression for \eqref{eq:Var_d_c} yields to the following decomposition
\begin{equation}
		\mathrm{Var}(H) = p_d \mathrm{Var}_d + p_c \mathrm{Var}_c + p_d p_c (\mu_d - \mu_c)^2.
\end{equation}
The QFI can therefore be decomposed as
\begin{equation}
	\mathcal{F}(T)=
	\mathcal{F}_{d}(T)
	+\mathcal{F}_{c}(T)
	+\mathcal{F}_{dc}(T),
\end{equation}
where
\begin{equation}
	\mathcal{F}_{d}=\frac{p_d\,\mathrm{Var}_d}{T^4},
	\quad
	\mathcal{F}_{c}=\frac{p_c\,\mathrm{Var}_c}{T^4},
	\quad
	\mathcal{F}_{dc}=
	\frac{p_d p_c(\mu_d-\mu_c)^2}{T^4}.
\end{equation}

In the low-temperature regime $T \to 0$ ($\beta \to \infty$), thermal occupation is strongly concentrated near the lowest accessible energies, while higher-energy contributions are exponentially suppressed. As a result, the continuum sector becomes negligible, yielding
\begin{equation}
    Z_c(\beta)\sim e^{-\beta E_c}\,\beta^{-(s+1)}\ll g_1e^{-\beta E_1}\qquad (\beta\to\infty).
\end{equation}
Since $E_c>E_1$, the continuum contribution decays faster than the first excited discrete level, and therefore
\begin{equation}
	Z(\beta) \simeq g_0 e^{-\beta E_0} + g_1 e^{-\beta E_1} \, .
\end{equation}	
Factoring out the ground state contribution yields
\begin{equation}
	Z(\beta) \simeq e^{-\beta E_0} \left( 1 + g_1 e^{-\beta \Delta} \right),
\end{equation}
where $\Delta = E_1 - E_0 > 0$. In this regime, the QFI takes the form
\begin{equation}
	\mathcal F(T) \simeq \frac{g_1 \Delta^2}{T^4} \exp\!\left(-\frac{\Delta}{T}\right).
\end{equation}
In the high-temperature regime $T \to \infty$ ($\beta \to 0$), the contribution of the discrete levels becomes negligible relative to the continuum sector, such that the partition function is asymptotically dominated by
\begin{equation}
	Z(\beta) \simeq \int_{E_c}^{\infty} \rho(E) e^{-\beta E} dE.
\end{equation}
Introducing $\epsilon = E - E_c \ge 0$, we obtain
\begin{equation}
	Z(\beta) \simeq e^{-\beta E_c} \int_0^{\infty} \rho(E_c + \epsilon) e^{-\beta \epsilon} d\epsilon.
\end{equation}
Given that we assume $\beta \to 0$, then $e^{-\beta E_c} \simeq 1$, and assuming that the continuum density of states has the high-energy asymptotic form $\rho(E_c+\epsilon)\sim C\epsilon^s$ as $\epsilon\to\infty$, we obtain
\begin{equation}
	Z(\beta) \simeq C \int_0^{\infty} \epsilon^s e^{-\beta \epsilon} d\epsilon = C \, \Gamma(s+1) \, \beta^{-(s+1)}.
\end{equation}
Here $\Gamma(z)$ is the Gamma function. From this, omitting the shift in the mean energy, we obtain
\begin{align}
	\langle H \rangle & = -\partial_\beta \ln Z = \frac{s+1}{\beta} = (s+1) T, \\
	\mathrm{Var}(H) & = \partial_\beta^2 \ln Z = \frac{s+1}{\beta^2} = (s+1) T^2.
\end{align}
Thus, the QFI behaves as
\begin{equation}
	\mathcal F(T) \simeq \frac{s+1}{T^2}.
\end{equation}
\bibliography{biblio.bib}

\end{document}